\documentclass[10pt]{revtex4}
\usepackage{amsmath}
\usepackage{amsfonts}
\usepackage{amssymb}
\usepackage{latexsym}
\usepackage{graphicx}
\graphicspath{{/}{figures/}}
\usepackage{epstopdf}
\begin{document}

\title{  Quarkonium in a thermal BIon }

\author{Alireza Sepehri $^{1}$}
\email{alireza.sepehri@uk.ac.ir} \affiliation{  $^{1}$ Research Institute for Astronomy and Astrophysics of
Maragha (RIAAM), P.O. Box 55134-441, Maragha, Iran. }

\author{Richard Pincak$^{2,3}$}\email{pincak@saske.sk}
\affiliation{ $^{2}$ Institute of Experimental Physics, Slovak Academy of Sciences,
Watsonova 47,043 53 Kosice, Slovak Republic}
\affiliation{ $^{3}$ Bogoliubov Laboratory of Theoretical Physics, Joint
Institute for Nuclear Research, 141980 Dubna, Moscow region, Russia}

\author{Michal \,Hnati\v{c} $^{4,5,6}$}\email{hnatic@saske.sk}
\affiliation{$^{4}$Faculty of Sciences, P.J. Safarik University, Ko\v{s}ice, Slovakia}
\affiliation{$^{5}$Institute of Experimental Physics, SAS,
         Watsonova 47, 040 01 Ko\v{s}ice, Slovakia
        }
\affiliation{$^{6}$Bogoliubov Laboratory of Theoretical Physics, Joint Institute for Nuclear Research,
        141 980 Dubna, Moscow Region, Russian Federation
        }

\author{Farook Rahaman $^{7}$}\email{rahaman@associates.iucaa.in}
\affiliation{$^{8}$Department of
Mathematics, Jadavpur University, Kolkata 700 032, West Bengal,
India.
        }

        \author{Anirudh Pradhan $^{8}$}\email{pradhan@associates.iucaa.in}

\affiliation{$^{9}$Department of
 Mathematics, Institute of Applied Sciences and Humanities, G L A
 University, Mathura-281 406, Uttar Pradesh, India.
        }

\begin{abstract}
 In the present article, the authors intend to propose a new theory which potentially allows the propagation of
 the formation and the evolution of quarkonium in a thermal BIon. When quarks are close to each other, quarkonium
 behaves like a scalar and by their getting away, it transits to a fermionic system. In order to analyze this
 particular behaviour, a new outlook approach needs to be adopted as the concurrent view is found deficient to
 analyse the aforesaid behaviour. Therefore, the authors' post deliberation accept the fermions and fermionic
 being cognate. We need to accept a theory that the origin of fermions and bosons be the same. However, in $M$-theory,
 these particles are independent and for this reason, \textbf{we use a new broader theory based on Lie-$N$-Algebra and we call
 it BLNA (Broad Lie-$N$-Algebra)} theory. Thus, the BLNA in a way the $M$-theory with $11$ dimensions.
 In this model, two types of energies with opposite signs emerge from nothing such as the sum over them
 becomes zero. They produce two types of branes with opposite quantum numbers and bosonic fields, which interact
 with each other and get compact. By compacting branes, the quarks and anti-quarks are produced on branes and exchange the
 graviton and the gravitino. These particles produce two types of wormholes which act opposite to each other. They preclude
 from closing or getting away of branes from each other and also occurrence of confinement. This confined potential
 which emerges from these wormholes depends on the separation distance between quarks and anti-quarks and also on
 temperature of system and is reduced to predicted potential in experiments and QCD. Also, total entropy of this
 system grows with increasing temperature and produces a repulsive force which leads to the separation of quarks
 and anti-quarks and also to the emergence of deconfinement.


 \end{abstract}
 \date{\today}

%%%%%%%%%%%%%%%%%%%%%%%%%%%%%%%%%%%%%%%%%%%%%%%%%%%%%%%%%%%%%%%%%%%%%%%%%%%%%%%%%%%%%%%%%%%%%%%%%%%%%%%%%%
%%%%%%%%%%%%%%%%%%%%%%%%%%%%%%%%%%%%%%%%% SECTION I %%%%%%%%%%%%%%%%%%%%%%%%%%%%%%%%%%%%%%%%%%%%%%%%%%%%%
\maketitle
\section{Introduction}
One of the main puzzles in QCD is the  reason for occurring confinement between quarks and anti-quarks which
prevents quarks from going much away from each other or coming very close to each other. Although, a transition
from hadronic matter to a plasma of deconfined quarks and gluons for high temperature can be observed \cite{k1}.
Many scientists have tried to resolve this problem. For example, in one investigation, authors have obtained the
quark-antiquark potential in $d$-dimensions by using the explicit solution of $d+1$-dimensional dilatonic gravity. Their
results were consistent with the predicted potential following IIB supergravity and Ads/CFT \cite{k2}. In another
research, authors have argued that quark-anti-quark potential depends on the curvature in IIB super-gravity background
with non-trivial dilaton and with curved four-dimensional space. This potential has shown  the appearance of the confinement
for the geometry type of hyperbolic (or de Sitter universe) and proposed the standard area law for large separations
\cite{k3}. In another paper, it has been discussed that the entropy of heavy quarkonium in strongly coupled quark-gluon
plasma,  grows with the inter-quark distance $r$ and thus, the  entropic force $F=T\frac{\partial S}{\partial r} $
($T$ is temperature) results in an anomalously strong quarkonium suppression in high temperatures. This means that for
high temperature, the entropic force leads to delocalization of the bound hadron states; and this delocalization can
be the mechanism indicating deconfinement \cite{k4}. Some other authors have discussed that color confinement leads
to formation of an event horizon for quarks and gluons which may be crossed only by quantum tunneling similar to black
holes. The radiation of this horizon is thermal; and its temperature is calculated by the chromodynamic force at the
confinement surface, and also it maintains color neutrality \cite{k5}. Recently, it has been shown
that color screening leads to an increase in the attractive binding force between the quarks, while the growing of
entropy with the separation between quarks and anti-quarks gives rise to a growing repulsion \cite{k6}. Motivated by
these research findings, we propose a new model which allows us to consider the origin of confining potential and
entropic force in BIonic system. BIon is a configuration of one brane and one anti-brane connected by one, two or
more wormholes \cite{k7,k8}. \\

To consider quarkonium, we need a BIon whose branes play the role of quarks and can't go very close or distant from
anti-branes which plays the role of anti-quarks. By closing of branes to each other, system behaves like
scalars and by getting away, it transits to two separated fermions. To construct this system, it is needed that the
origin of fermions and bosons be the same; and they can be changed to each other. In $M$-theory, fermions are completely
independent of bosons and for this reason, we propose the new BLNA theory where the number of its degrees of freedom
and dimensions is more than $M$-theory and is reduced to $11$ dimensions. In this model, first, there is nothing with
zero energy and degrees of freedom. Then, two energies, one with positive sign and other with negative sign are produced and
each of them has its own degrees of freedom. These degrees of freedoms produce two branes with opposite quantum numbers
which are known as branes and anti-branes. On these branes, only bosons like scalars and gravitons live which interact with
fields of other branes and lead to the compacting of branes. The compacting of branes produces fermions like quarks,
antiquarks and gravitino. Quarks are placed on branes and exchange gravitons and gravitinos with anti-quarks that
are located on anti-branes. Gravitons produce one bosonic wormhole subject to distance between quarks and
antiquarks, leads to the emergence of attractive force and confinement between these particles  and prevents them getting
away from each other. Gravitino creates the fermionic wormhole which causes the production of repulsive force
in small separation distance between quarks and anti-quarks and prevents the proximity to each other. These results are
in agreement with predictions of QCD that quarks and anti-quarks can't go much away or much close to each other.
By increasing temperature, the entropy of bosonic wormhole increases and one of fermionic wormhole decreases and one repulsive
force emerges which leads to deconfinement. \\

According to Nojiri and Odintsov \cite{nn1,nn2}, there is some ground to believe that gravitational alternative to dark energy
(which may be called the effective dark energy) may be produced by the modification of General Relativity which is dictated
by string/$M$-theory. Modified gravity presents very natural modification of the early-time inflation and late time acceleration.
The authors \cite{nn1,nn2} concluded that such a model which seems to eliminate the need of dark energy may have the origin in
$M$-theory. In $11$-dimensional $M$-theory and $10$-dimensional string theory, there are many puzzles and questions remain
unanswered. The limitations of the aforesaid framework create an extensive scope for researchers to work for rather more
encompassing theory. For example, in $M$-theory, only there are two stable objects, $M2$ and $M5$. While for constructing
four-dimensional universe, we need to $M3$-brane which is unstable in it. On the other hand, for both $M2$ and $M5$-branes,
a lie-$3$-algebra has been suggested (see for example, \cite{n1,n2}). Extending this algebra to N-dimensions and dimensions
from $11$ to $M$, we search for a bigger theory that solves the puzzles in $M$-theory. Additionally, the origin of potentials and
confinement in QCD is also unclear. Specially, strong coupling constant is a free parameter and its value can be determined
through experiments. Attempts have been made by various scientists to locate its origin in the string theory and also to obtain
the value of this constant in this theory which be in accordance with experiments, however no substantial progress could be made
in the direction, owing to the limitations of the concurrent propounded theories. In our proposed theory, all parameters of QCD
can be determined without adding any value manually. Also, our theory explains the origin of confinement. \\

In section \ref{o1} of the article, the process of formation of BIon in BLNA-theory and construct quarkonium on it is considered, while
in section \ref{o2}, the relation between potential and entropy of quarkonium with temperature in BIon is studied.
The last section is devoted to the summary and conclusion. \\

Units throughout the paper are: $\hbar=c=8\pi G=1$.

%%%%%%%%%%%%%%%%%%%%%%%%%%%%%%%%%%%%%%%%%%%%%%%%%%%%%%%%%%%%%%%%%%%%%%%%%%%%%%%%%%%%%%%%%%%%%%%%%%%%%%%%%%%%%%
%%%%%%%%%%%%%%%%%%%%%%%%%%%%%%%%%%%%%%%%%%% SECTION II %%%%%%%%%%%%%%%%%%%%%%%%%%%%%%%%%%%%%%%%%%%%%%%%%%%%%%
\section{Emergence of quarkonium in BIonic system  }\label{o1}
In this section, the process of formation of a BIon in Lie-$N$-algebra is considered in the beginning itself. We  assume that there is
a null at first. Then, two positive and negative energies emerge such that the sum over them becomes zero. After
that, these energies get excited and produce a brane and an anti-brane. Bosons which live on these branes, interact
with those on anti-branes and produce a bosonic wormhole between two branes. Also, fermions which are placed on these
branes create a fermionic wormhole. This system behaves like a BIon with two separate wormholes which may act reverse
to each other. We will construct the quarkonium in this system and extract the predicted potential in QCD. \\

To construct BIonic system in Lie-$N$-algebra, first, we should extend the usual actions in string theory and $M$-theory
to a world with $M$ dimensions and Lie-$N$-algebra. Previously, it has been shown that all Dp-branes in string theories are
constructed from $D0$-branes \cite{k7,k8,k9,k10,k11,k12,k13,k14,k15,k16}. Also, all $Mp$-branes in $M$-theory are built from
$M0$-branes \cite{k17,k18,k19,k20}. The difference between $D0$-branes and $M0$-branes is in the dimensions of their actions and
their algebra. The action of $M0$-branes contains three dimensional brackets which obey rules of Lie-three-algebra, while the
action of $D0$-branes has two dimensional brackets which obey that of Li-two-algebra. In these theories, by joining $M0$/$D0$-branes
and formation of higher dimensional branes, gauge fields emerge and by compacting branes-anti-symmetrically, fermions
emerge. Now, we extend these theories by increasing dimensions of brackets in action of $M0$ and $D0$ to $N$ and using of
Lie-$N$-algebra. We name this new theory as BLNA-theory. In BLNA-theory, we can show that the origin of bosons and fermions
are the same and for this reason, it can be applied for considering the behaviour of some systems like quarkonium. In
addition to this, BLNA-theory can be reduced to $M$-theory, only by putting $N=3$ and to string theory, by putting $N=2$. \\

First, we will show that the action of $Dp$-branes can be constructed by multiplying the action of $D0$-branes. Then, we
generalize this mechanism to eleven dimensional $M$-theory and calculate the action of $Mp$-branes by multiplying the
action of $M0$-branes. The action for $D1$-brane is \cite{k7,k8,k9,k10,k11,k12,k13,k14,k15,k16}:

\begin{eqnarray}
S = - T_{D1}\int  d^{2}\sigma ~ STr \Biggl(-det(P_{ab}[E_{mn}
E_{mi}(Q^{-1}+\delta)^{ij}E_{jn}]+ \lambda F_{ab})det(Q^{i}_{j})\Biggr)^{1/2},
\label{r1}
\end{eqnarray}

where

\begin{eqnarray}
   E_{mn} = G_{mn} + B_{mn}, \qquad  Q^{i}_{j} = \delta^{i}_{j} + i\lambda[X^{j},X^{k}]E_{kj} \label{r2}
\end{eqnarray}

 where $\lambda=2\pi l_{s}^{2}$, $G_{ab}=\eta_{ab}+\partial_{a}X^{i}\partial_{b}X^{i}$ and $X^{i}$ are attached scalar strings
 to branes. In this relation, $a,b=0,1,...,p$ refer to the world-volume indices of the Dp-branes, $i,j,k = p+1,...,9$
denote indices of the transverse space, and $m$,$n$ are corresponded to
ten-dimensional spacetime indices. Also,
$T_{Dp}=\frac{1}{g_{s}(2\pi)^{p}l_{s}^{p+1}}$ denotes the tension of
$Dp$-brane, $l_{s}$ refers to the string length and $g_{s}$ is the string
coupling. To obtain the action for $Dp$-brane, we should use
the below relations \cite{k7,k8,k9,k10,k11,k12,k13,k14,k15,k16}:

\begin{eqnarray}
&& \Sigma_{a=0}^{p}\Sigma_{m=0}^{9}\rightarrow \frac{1}{(2\pi l_{s})^{p}}\int d^{p+1}\sigma \Sigma_{m=p+1}^{9}\Sigma_{a=0}^{p}
\qquad \lambda = 2\pi l_{s}^{2}\nonumber \\
&& i,j=p+1,..,9\qquad a,b=0,1,...p\qquad m,n=0,1,..,9 \nonumber \\
&& i,j\rightarrow a,b \Rightarrow [X^{a},X^{i}]=i \lambda
\partial_{a}X^{i}\qquad  [X^{a},X^{b}]=\frac{ i \lambda F^{ab}}{2} \nonumber \\
&& \frac{1}{Q}\rightarrow \sum_{n=1}^{p}
\frac{1}{Q}(\partial_{a}X^{i}\partial_{b}X^{i}+\frac{\lambda^{2}}{4} (F^{ab})^{2})^{n}\nonumber \\
&& det(Q^{i}_{j})\rightarrow det(Q^{i}_{j})\prod_{n=1}^{p}
det(\partial_{a_{n}}X^{i}\partial_{b_{n}}X^{i}+\frac{\lambda^{2}}{4} (F^{a_{n}b_{n}})^{2}) \label{r3}
\end{eqnarray}

Applying above relations in action given by (\ref{r1}), we obtain
the action for $Dp$-brane \cite{k7,k8,k9,k10,k11,k12,k13,k14,k15,k16,ss20}:

\begin{eqnarray}
 &&S=-\frac{T_{Dp}}{2}\int d^{p+1}x \sum_{n=1}^{p}\beta_{n} \chi^{\mu_{0}}_{[\mu_{0}}
 \chi^{\mu_{1}}_{\mu_{1}}...\chi^{\mu_{n}}_{\mu_{n}]},\label{s4}
\end{eqnarray}

where

\begin{eqnarray}
 &&\chi^{\mu}_{\nu}\equiv
\sqrt{g^{\mu\rho}\partial_{\rho}X^{i}\partial_{\nu}X^{j}\eta_{ij}+\frac{\lambda^{2}}{4} (F^{ab})^{2}
}\label{s5}
\end{eqnarray}

and $X^{a}$'s refer to scalars, $\mu,\nu=0,1,...,p$ denote to the
world-volume indices of the $Mp$-branes, $i,j,k = p+1,...,9$ refer to
 indices of the transverse space and $\beta$ is a constant.
Also, $T_{Dp}=\frac{1}{g_{s}(2\pi)^{p}l_{s}^{p+1}}$ denotes the tension
of $Dp$-brane, $l_{s}$ refers to the string length and $g_{s}$ is the
string coupling. Now, we can assert that this action can be built
by multiplying over the actions for D0-branes. Applying the mechanism in ref
\cite{k8}, we can obtain the below relations \cite{k7,k8,k9,k10,k11,k12,k13,k14,k15,k16}:

\begin{eqnarray}
 && [X_{\rho},X^{i}]=i\lambda \partial_{\rho}X^{i}\qquad  [X^{a},X^{b}]=
 \frac{ i \lambda F^{ab}}{2}\rightarrow \chi^{\mu}_{\nu}\equiv
\sqrt{g^{\mu\rho}[X_{\rho},X^{m}][X_{\nu},X^{n}]\eta_{mn}}
\nonumber\\&& \Sigma_{m=0}^{9} \rightarrow
\Sigma_{a,b=0}^{p}\Sigma_{j=p+1}^{9}\label{s6}
\end{eqnarray}

Using (\ref{s6}) in (\ref{s5}), we obtain:

\begin{eqnarray}
&& S_{Dp} = -(T_{D0})^{p} \int dt \sum_{n=1}^{p}\beta_{n}\Big(
\delta^{a_{1},a_{2}...a_{n}}_{b_{1}b_{2}....b_{n}}L^{b_{1}}_{a_{1}}...L^{b_{n}}_{a_{n}}\Big)^{1/2}\nonumber\\&&
(L)^{b}_{a}=Tr\Big( \Sigma_{a,b=0}^{p}\Sigma_{j=p+1}^{9}(
[X^{b},X^{j}][X_{a},X_{j}]+[X^{b},X^{b'}][X_{a},X_{b'}]+[X^{i},X^{j}][X_{i},X_{j}]\delta^{b}_{a})\Big), \label{s7}
\end{eqnarray}

where we have defined the antisymmetric properties for $\delta$ and
used of the action of $D0$-brane \cite{k7,k8,k9,k10,k11,k12,k13,k14,k15,k16}:

\begin{eqnarray}
&& S_{D0} = -T_{D0} \int dt Tr( \Sigma_{m=0}^{9}
[X^{m},X^{n}]^{2}) \label{s8}
\end{eqnarray}

Equation (\ref{s7}) indicates that each $Dp$-brane can be built from linking $p$ $D0$-branes. This
mechanism can be applied in $M$-theory, and each $Mp$-brane can be constructed by joining $M0$-branes.
By substituting three dimensional Nambu-Poisson bracket for Mp-branes instead of two one in action and using
the Li-$3$-algebra \cite{k17,k18,k19,k20}, we can obtain the relevant action for $M0$-brane
\cite{k7,k8,k9,k10,k11,k12,k13,k14,k15,k16,k17,k18,k19,k20}:

\begin{eqnarray}
S_{M0} = T_{M0}\int dt Tr\left( \Sigma_{M,N,L=0}^{10}
\langle[X^{M},X^{N},X^{L}],[X^{M},X^{N},X^{L}]\rangle\right), \label{s9}
\end{eqnarray}

where $X^{M}=X^{M}_{\alpha}T^{\alpha}$ and

\begin{eqnarray}
 &&[T^{\alpha}, T^{\beta}, T^{\gamma}]= f^{\alpha \beta \gamma}_{\eta}T^{\eta} \nonumber \\&&
 \langle T^{\alpha}, T^{\beta} \rangle = h^{\alpha\beta} \nonumber \\&& [X^{M},X^{N},X^{L}]=
 [X^{M}_{\alpha}T^{\alpha},X^{N}_{\beta}T^{\beta},X^{L}_{\gamma}T^{\gamma}]\nonumber \\&&
 \langle X^{M},X^{M}\rangle = X^{M}_{\alpha}X^{M}_{\beta}\langle T^{\alpha}, T^{\beta} \rangle,
\label{s10}
\end{eqnarray}

where  $X^{M}$($i=1,3,...10$) refer to  scalars which are attached to M0-brane.
Replacing the action of $D0$ by $M0$ in the action (\ref{s7}), we get
\cite{k7,k8,k9,k10,k11,k12,k13,k14,k15,k16,k17,k18,k19,k20}:

\begin{eqnarray}
&&S_{Mp} = -(T_{M0})^{p} \int dt \sum_{n=1}^{p}\beta_{n}\Big(
\delta^{a_{1},a_{2}...a_{n}}_{b_{1}b_{2}....b_{n}}L^{b_{1}}_{a_{1}}...L^{b_{n}}_{a_{n}}\Big)^{1/2}\nonumber\\&&
(L)^{a}_{b}= Tr\Big(  \Sigma_{a,b=0}^{p}\Sigma_{j=p+1}^{10}(
\langle[X^{a},X^{i},X^{j}],[X_{b},X_{i},X_{j}]\rangle+\langle[X^{a},X^{c},X^{j}],[X_{b},X_{c},X_{j}]
\rangle+\nonumber\\&&
\langle[X^{a},X^{c'},X^{c}],[X_{b},X_{c'},X_{c}]\rangle+\langle[X^{k},X^{i},X^{j}],[X_{k},X_{i},X_{j}]
\rangle\delta^{a}_{b})\Big) \label{s11}
\end{eqnarray}

Following the mechanism for $Dp$-branes in string theory, different $Mp$-branes can be constructed from
$M0$-brane by applying the following mappings \cite{k8,k9,k10,k11,k12,k13,k14,k15,k16,k17,k18,k19,k20}:

\begin{eqnarray}
&&\langle[X^{a},X^{b},X^{i}],[X^{a},X^{b},X^{i}]\rangle=
 \frac{1}{2}\langle \partial_{b}\partial_{a}X^{i},\partial_{b}\partial_{a}X^{i}\rangle \nonumber \\
&& \langle[X^{i},X^{b},X^{i}],[X^{i},X^{b},X^{i}]\rangle=
 \frac{1}{2}\sum_{j}(X^{j})^{2}\langle \partial_{b}X^{i},\partial_{b}X^{i}\rangle\nonumber \\
&&\nonumber \\
&&\langle[X^{a},X^{b},X^{c}],[X^{a},X^{b},X^{c}]\rangle=\frac{\lambda^{2}}{6}(F^{abc}_{\alpha\beta\gamma})
(F^{abc}_{\alpha\beta\eta})\langle[T^{\alpha},T^{\beta},T^{\gamma}],[T^{\alpha},T^{\beta},T^{\eta}]\rangle)=\nonumber \\ &&
\frac{\lambda^{2}}{6}(F^{abc}_{\alpha\beta\gamma})(F^{abc}_{\alpha\beta\eta})f^{\alpha \beta \gamma}_{\sigma}
h^{\sigma \kappa}f^{\alpha \beta \eta}_{\kappa} \langle T^{\gamma},T^{\eta}\rangle=\frac{\lambda^{2}}{6}
(F^{abc}_{\alpha\beta\gamma})(F^{abc}_{\alpha\beta\eta})\delta^{\kappa \sigma} \langle T^{\gamma},T^{\eta}\rangle=
\frac{\lambda^{2}}{6}\langle F^{abc},F^{abc}\rangle\nonumber \\
&&\langle[X^{i},X^{b},X^{c}],[X^{i},X^{b},X^{c}]\rangle=\frac{\lambda^{2}}{4}\sum_{j}(X^{j})^{2}
\langle F^{bc},F^{bc}\rangle\nonumber \\
&&\nonumber \\
&&\Sigma_{m}\rightarrow \frac{1}{(2\pi)^{p}}\int d^{p+1}\sigma \Sigma_{m-p-1}
i,j=p+1,..,10\quad a,b=0,1,...p\quad m,n=0,..,10~~
\label{a31}
\end{eqnarray}

where

\begin{eqnarray}
&&F_{abc}=\partial_{a} A_{bc}-\partial_{b} A_{ca}+\partial_{c} A_{ab}\label{a32}
\end{eqnarray}

and $A_{ab}$ is $2$-form gauge field. Using the mappings of Eq. (\ref{a31})
in action (\ref{s11}), we can obtain the relevant action for $Mp$-brane

\begin{eqnarray}
&&S_{Mp} = -(T_{M0})^{p} \int dt \sum_{n=1}^{p}\beta_{n}\Big(
\delta^{a_{1},a_{2}...a_{n}}_{b_{1}b_{2}....b_{n}}L^{b_{1}}_{a_{1}}...L^{b_{n}}_{a_{n}}\Big)^{1/2}\nonumber\\&&
(L)^{a}_{b}= \delta^{a}_{b}Tr
(\Sigma_{a,b,c=0}^{p}
\Sigma_{i,j,k=p+1}^{10}
\{\frac{1}{2}\sum_{j}(X^{j})^{2}\langle\partial_{a}X^{i},\partial_{a}X^{i}\rangle
+\frac{1}{2}\langle \partial_{b}\partial_{a}X^{i},\partial_{b}\partial_{a}X^{i}\rangle+\nonumber \\ &&\frac{\lambda^{2}}{6}
\langle F_{abc},F_{abc}\rangle+\frac{\lambda^{2}}{4}\sum_{j}(X^{j})^{2}
\langle F_{bc},F_{bc}\rangle  -\frac{1}{4}\langle[X^{i},X^{j},X^{k}],[X^{i},X^{j},X^{k}]\rangle
\})\label{a33}
\end{eqnarray}

 This action is in good agreement with previous predictions of M-theory \cite{k7,k8,k9,k10,k11,k12,k13,k14,k15,k16,k17,k18,k19,k20}.
 In this theory, rank of fields change from zero to $2$, which rank zero is related to scalar ($X$), rank one is the vector ($A^{a}$)
 and rank two is corresponded to tensor fields ($A^{ab}$) like gravitons.

  Now, by replacing the brackets in actions of (\ref{s8} and \ref{s9}) by $N$-dimensional brackets and extending dimensions
  to $M$, we define the action of $G0$-brane as:

 \begin{eqnarray}
 S_{G0} = T_{G0}\int dt Tr\left( \Sigma_{L_{1}=L_{2}..L_{N}=0}^{M}
 \langle[X^{L_{1}},X^{L_{2}},...X^{L_{N}}],[X^{L_{1}},X^{L_{2}},...X^{L_{N}}]\rangle\right), \label{P1}
 \end{eqnarray}

 where $X^{M}=X^{M}_{\alpha}T^{\alpha}$ and

 \begin{eqnarray}
  &&[T^{\alpha_{1}}, T^{\alpha_{2}}..T^{\alpha_{N}}]= f^{\alpha_{1}..\alpha_{N}}_{\alpha_{L}}T^{L} \nonumber \\&&
  \langle T^{\alpha}, T^{\beta} \rangle = h^{\alpha\beta} \nonumber \\&& [X^{L_{1}},X^{L_{2}},...X^{L_{N}}]=
  [X^{L_{1}}_{\alpha_{1}}T^{\alpha_{1}},X^{L_{2}}_{\alpha_{2}}T^{\alpha_{2}},...X^{L_{N}}_{\alpha_{N}}T^{\alpha_{N}}]
  \nonumber \\&&\langle X^{M},X^{M}\rangle = X^{M}_{\alpha}X^{M}_{\beta}\langle T^{\alpha}, T^{\beta} \rangle
 \label{P2}
 \end{eqnarray}

  This action can be reduced to the action of $M0$-branes by puting $N=3$ and $M=10$ and reduced to the action of $D0$-brane
  for $N=2$ and $M=9$. By replacing three dimensional brackets with $N$-dimensional brackets and increasing dimensions from
  $11$ to $M$ in action (\ref{s11}), we can calculate the action of $Gp$-brane:

  \begin{eqnarray}
  &&S_{Gp} = -(T_{G0})^{p} \int dt \sum_{n=1}^{p}\beta_{n}\Bigl(
  \delta^{a_{1},a_{2}...a_{n}}_{b_{1}b_{2}....b_{n}}L^{b_{1}}_{a_{1}}...L^{b_{n}}_{a_{n}}\Bigr)^{1/2}\nonumber\\&&
  (L)^{a_{n}}_{b_{n}}= \delta^{a_{n}}_{b_{n}}Tr\Bigl( \Sigma_{L=0}^{N} \Sigma_{H=0}^{N-L}\Sigma_{a_{1}..a_{L}=0}^{p}
  \Sigma_{j_{1}..j_{H}=p+1}^{M}( \langle[X^{j_{1}},..X^{j_{H-1}},X^{a_{1}},..X^{a_{L}},X^{j_{H}}],
  \langle[X^{j_{1}},..X^{j_{H-1}},X^{a_{1}},..X^{a_{L}},X^{j_{H}}]\rangle) + \nonumber \\
         && \Sigma_{L=0}^{N} \Sigma_{H=0}^{N-L}\Sigma_{a_{1}..a_{L}=0}^{p}\Sigma_{j_{1}..j_{H}=p+1}^{M}
              (\langle[X^{j_{1}},..X^{j_{H}},X^{a_{1}},..X^{a_{L}}],[X^{j_{1}},..X^{j_{H}},X^{a_{1}},..X^{a_{L}}]
              \rangle) \Bigr) \label{P3}
  \end{eqnarray}

  Generalizing the laws given by Eq. (\ref{a31}) for $M$-theory to $N$-dimensional brackets in BLNA-theory, we can write
  following mappings:

  \begin{eqnarray}
  && \Sigma_{L=0}^{N} \Sigma_{H=0}^{N-L}\Sigma_{a_{1}..a_{L}=0}^{p}\Sigma_{j_{1}..j_{H}=p+1}^{M}
  \langle[X^{j_{1}},..X^{j_{H-1}},X^{a_{1}},..X^{a_{L}},X^{j_{H}}],\langle[X^{j_{1}},..X^{j_{H-1}},X^{a_{1}},..X^{a_{L}},X^{j_{H}}]
  \rangle=\nonumber \\
       &&
   \frac{1}{2}\Sigma_{L=0}^{N} \Sigma_{H=0}^{N-L}\Sigma_{a_{1}..a_{L}=0}^{p}\Sigma_{j_{1}..j_{H}=p+1}^{M}(X^{j_{1}}..X^{j_{H-1}})^{2}
   \langle \partial_{a_{1}}..\partial_{a_{L}}X^{i},\partial_{a_{1}}..\partial_{a_{L}}X^{i}\rangle\nonumber \\
     &&\nonumber \\
       &&\nonumber \\
  && \Sigma_{L=0}^{N} \Sigma_{H=0}^{N-L}\Sigma_{a_{1}..a_{L}=0}^{p}\Sigma_{j_{1}..j_{H}=p+1}^{M}
     \langle[X^{j_{1}},..X^{j_{H}},X^{a_{1}},..X^{a_{L}}],[X^{j_{1}},..X^{j_{H}},X^{a_{1}},..X^{a_{L}}]\rangle=\nonumber \\
         &&
        \Sigma_{L=0}^{N} \Sigma_{H=0}^{N-L}\Sigma_{a_{1}..a_{L}=0}^{p}\Sigma_{j_{1}..j_{H}=p+1}^{M}
        \frac{\lambda^{2}}{1.2...N}(X^{j_{1}}..X^{j_{H}})^{2}\langle F^{a_{1}..a_{L}}, F^{a_{1}..a_{L}}\rangle\nonumber \\
  &&\nonumber \\
    &&\nonumber \\
      &&\nonumber \\
        &&F_{a_{1}..a_{n}}=\partial_{[a_{1}} A_{a_{2}..a_{n}]}=\partial_{a_{1}} A_{a_{2}..a_{n}}-
        \partial_{a_{2}} A_{a_{1}..a_{n}}+..\nonumber \\
          &&\nonumber \\
            &&\nonumber \\
  &&\Sigma_{m}\rightarrow \frac{1}{(2\pi)^{p}}\int d^{p+1}\sigma \Sigma_{m-p-1}
  i,j=p+1,..,M\quad a,b=0,1,...p\quad m,n=0,..,M~~
  \label{P4}
  \end{eqnarray}

    Substituting mappings of Eq. (\ref{P4}) in action (\ref{P3}), we obtain the following action for $Gp$-branes:

     \begin{eqnarray}
     &&S_{Gp} = -(T_{Gp}) \int dt \sum_{n=1}^{p}\beta_{n}\Big(
     \delta^{a_{1},a_{2}...a_{n}}_{b_{1}b_{2}....b_{n}}L^{b_{1}}_{a_{1}}...L^{b_{n}}_{a_{n}}\Big)^{1/2}\nonumber\\&&
     (L)^{a_{n}}_{b_{n}}= \delta^{a_{n}}_{b_{n}}Tr\Big(
        \frac{1}{2}\Sigma_{L=0}^{N} \Sigma_{H=0}^{N-L}\Sigma_{a_{1}..a_{L}=0}^{p}
        \Sigma_{j_{1}..j_{H}=p+1}^{M}(X^{j_{1}}..X^{j_{H-1}})^{2}\langle \partial_{a_{1}}..\partial_{a_{L}}X^{i},
        \partial_{a_{1}}..\partial_{a_{L}}X^{i}\rangle + \nonumber \\
            &&
          \Sigma_{L=0}^{N} \Sigma_{H=0}^{N-L}\Sigma_{a_{1}..a_{L}=0}^{p}\Sigma_{j_{1}..j_{H}=p+1}^{M}
         \frac{\lambda^{2}}{1.2...N}(X^{j_{1}}..X^{j_{H}})^{2}\langle F^{a_{1}..a_{L}}, F^{a_{1}..a_{L}}\rangle \Big) \label{P5}
     \end{eqnarray}

   This action is reduced to action of $Dp$-brane (\ref{s7}) for $N=2$ and $M=9$ and action of $Mp$-brane (\ref{a33}) for
   $N=3$ and $M=10$. This action is not complete, because, we have ignored fermions in it. In fact, in cosmology and
   other systems like quarkonium, we need supersymmetry and in order for it to be produced, we need certain degrees
   of freedom for bosons and fermions to be the same. Previously, it has been shown that by compacting part of brane,
   the symmetry of system is broken and scalars decay to fermions ($X \rightarrow \psi^{U}\psi^{L}$ ) \cite{k9}.
   We use the mechanism in \cite{k9}, and compactify $M^{th}$ dimension of branes on a circle with radius $R$ by
   choosing  $<X^{M}>=i\frac{R}{l_{p}^{1/2}}T^{M}$ for boson and $<\psi^{L,M}>=i\frac{R^{1/2}}{l_{p}^{1/4}}T^{L,M}$
   for fermions in action of (\ref{P1}). We obtain the following action for $G0$-brane:

   \begin{eqnarray}
 &&   S_{G0} = S_{G0,non-compact}+S_{G0,compact} =\nonumber \\
               &&  T_{G0}\int dt Tr \Big( \Sigma_{L_{1}=L_{2}..L_{N}=0}^{M}
  ( \langle[X^{L_{1}},X^{L_{2}},...X^{L_{N}}],[X^{L_{1}},X^{L_{2}},...X^{L_{N}}]\rangle \nonumber \\
              &&
  - i\frac{R^{2}}{l_{p}}\langle[T^{L_{1}},X^{L_{2}},...\psi^{R,L_{N}}],[X^{L_{1}},X^{L_{2}},...\psi^{R,L_{N}}]\rangle )\Big) \label{P6}
   \end{eqnarray}

We can choose $\gamma^{L_{1}}=T^{L_{1}}\frac{R^{2}}{l_{p}}$ where $\gamma^{L_{1}}$'s are the Pauli matrices in $M$ dimensions
and rewrite action of $G0$-brane as follows:

   \begin{eqnarray}
 &&   S_{G0} = T_{G0}\int dt Tr \Big( \Sigma_{L_{1}=L_{2}..L_{N}=0}^{M}
  ( \langle[X^{L_{1}},X^{L_{2}},...X^{L_{N}}],[X^{L_{1}},X^{L_{2}},...X^{L_{N}}]\rangle \nonumber \\
              &&
        - i\langle[\gamma^{L_{1}},X^{L_{2}},...\psi^{R,L_{N}}],[X^{L_{1}},X^{L_{2}},...\psi^{R,L_{N}}]\rangle )\Big) \label{P7}
   \end{eqnarray}

 It is clear that brackets in this action include both degrees of freedoms for bosons and fermions and generators of algebra
 behave like the Pauli matrices in $M$ dimensions. Replacing these brackets with brackets of Eq. (\ref{P3}), we can
 calculate the action of $Gp$-brane:

   \begin{eqnarray}
   &&S_{Gp} = -(T_{G0})^{p} \int dt \sum_{n=1}^{p}\beta_{n}\Big(
   \delta^{a_{1},a_{2}...a_{n}}_{b_{1}b_{2}....b_{n}}L^{b_{1}}_{a_{1}}...L^{b_{n}}_{a_{n}}\Big)^{1/2}\nonumber\\&&
   (L)^{a_{n}}_{b_{n}}= \delta^{a_{n}}_{b_{n}}Tr\Big(\Sigma_{L=0}^{N} \Sigma_{H=0}^{N-L}\Sigma_{a_{1}..a_{L}=0}^{p}
   \Sigma_{j_{1}..j_{H}=p+1}^{M}( \langle[X^{j_{1}},..X^{j_{H-1}},X^{a_{1}},..X^{a_{L}},X^{j_{H}}],
   \langle[X^{j_{1}},..X^{j_{H-1}},X^{a_{1}},..X^{a_{L}},X^{j_{H}}]\rangle) + \nonumber \\
     && \Sigma_{L=0}^{N} \Sigma_{H=0}^{N-L}\Sigma_{a_{1}..a_{L}=0}^{p}\Sigma_{j_{1}..j_{H}=p+1}^{M}
        (\langle[X^{j_{1}},..X^{j_{H}},X^{a_{1}},..X^{a_{L}}],[X^{j_{1}},..X^{j_{H}},X^{a_{1}},..X^{a_{L}}]\rangle- \nonumber \\
    &&  i\langle[\gamma^{j_{1}},..X^{j_{H-1}},X^{a_{1}},..X^{a_{L}},\psi^{R,j_{H}}],\langle[X^{j_{1}},..X^{j_{H-1}},
    X^{a_{1}},..X^{a_{L}},\psi^{R,j_{H}}]\rangle) - \nonumber \\
      && i\Sigma_{L=0}^{N} \Sigma_{H=0}^{N-L}\Sigma_{a_{1}..a_{L}=0}^{p}\Sigma_{j_{1}..j_{H}=p+1}^{M}
 (\langle[\gamma^{j_{1}},..\psi^{R,j_{H}},X^{a_{1}},..X^{a_{L}}],[X^{j_{1}},..\psi^{R,j_{H}},X^{a_{1}},..X^{a_{L}}]\rangle)\Big)
 \label{P8}
   \end{eqnarray}

To obtain the general form of action in terms of spinor fields, we add following laws to equation (\ref{P4})

  \begin{eqnarray}
  && \Sigma_{L=0}^{N} \Sigma_{H=0}^{N-L}\Sigma_{a_{1}..a_{L}=0}^{p}\Sigma_{j_{1}..j_{H}=p+1}^{M}
  \langle[\gamma^{j_{1}},..X^{j_{H-1}},X^{a_{1}},..X^{a_{L}},\psi^{j_{H}}],\langle[X^{j_{1}},..X^{j_{H-1}},X^{a_{1}},..X^{a_{L}},
  \psi^{j_{H}}]\rangle=\nonumber \\
       &&
   \frac{1}{2}i\Sigma_{L=0}^{N} \Sigma_{H=0}^{N-L}\Sigma_{a_{1}..a_{L}=0}^{p}\Sigma_{j_{1}..j_{H}=p+1}^{M}
   (X^{j_{1}}..X^{j_{H-1}})^{2}\gamma^{a_{L-1}}\langle \partial_{a_{1}}..\partial_{a_{L-1}}\psi^{i},\partial_{a_{1}}..
   \partial_{a_{L}}\psi^{i}\rangle\nonumber \\
     &&\nonumber \\
       &&\nonumber \\
  && \Sigma_{L=0}^{N} \Sigma_{H=0}^{N-L}\Sigma_{a_{1}..a_{L}=0}^{p}\Sigma_{j_{1}..j_{H}=p+1}^{M}
     \langle[X^{j_{1}},..X^{j_{H}},X^{a_{1}},..X^{a_{L}}],[X^{j_{1}},..X^{j_{H}},X^{a_{1}},..X^{a_{L}}]\rangle=\nonumber \\
         &&
        i\Sigma_{L=0}^{N} \Sigma_{H=0}^{N-L}\Sigma_{a_{1}..a_{L}=0}^{p}\Sigma_{j_{1}..j_{H}=p+1}^{M}\frac{\lambda^{2}}{1.2...N}
   (X^{j_{1}}..X^{j_{H}})^{2}\gamma^{a_{L-1}}\langle \bar{F}^{a_{1}..a_{L-1}}, \bar{F}^{a_{1}..a_{L}}\rangle\nonumber \\
  &&\nonumber \\
    &&\nonumber \\
      &&\nonumber \\
        &&\bar{F}_{a_{1}..a_{n}}=\partial_{[a_{1}} \bar{A}_{a_{2}..a_{n}]}=\partial_{a_{1}} \bar{A}_{a_{2}..a_{n}}-
        \partial_{a_{2}} \bar{A}_{a_{1}..a_{n}}+..\nonumber \\
          &&\nonumber \\
            &&\nonumber \\
  &&\Sigma_{m}\rightarrow \frac{1}{(2\pi)^{p}}\int d^{p+1}\sigma \Sigma_{m-p-1}
  i,j=p+1,..,M\quad a,b=0,1,...p\quad m,n=0,..,M~~
  \label{P9}
  \end{eqnarray}

  where $\bar{A}_{a_{2}..a_{n}}$ are fermionic superpartners of gauge bosons $A_{a_{2}..a_{n}}$ and $\psi$
  are the fermionic superpartner of scalar strings $X$. Replacing rules of Eq. (\ref{P9}) in action of (\ref{P8}), we get:

       \begin{eqnarray}
       &&S_{Gp} = -(T_{Gp}) \int dt \sum_{n=1}^{p}\beta_{n}\Big(
       \delta^{a_{1},a_{2}...a_{n}}_{b_{1}b_{2}....b_{n}}L^{b_{1}}_{a_{1}}...L^{b_{n}}_{a_{n}}\Big)^{1/2}\nonumber\\&&
       (L)^{a_{n}}_{b_{n}}= \delta^{a_{n}}_{b_{n}}Tr\Big(\frac{1}{2}\Sigma_{L=0}^{N} \Sigma_{H=0}^{N-L}
       \Sigma_{a_{1}..a_{L}=0}^{p}\Sigma_{j_{1}..j_{H}=p+1}^{M}(X^{j_{1}}..X^{j_{H-1}})^{2}\langle \partial_{a_{1}}..
       \partial_{a_{L}}X^{i},\partial_{a_{1}}..\partial_{a_{L}}X^{i}\rangle + \nonumber \\
              &&
      \Sigma_{L=0}^{N} \Sigma_{H=0}^{N-L}\Sigma_{a_{1}..a_{L}=0}^{p}\Sigma_{j_{1}..j_{H}=p+1}^{M}
      \frac{\lambda^{2}}{1.2...N}(X^{j_{1}}..X^{j_{H}})^{2}\langle F^{a_{1}..a_{L}}, F^{a_{1}..a_{L}}\rangle  -\nonumber\\ &&
       \frac{1}{2}i\Sigma_{L=0}^{N} \Sigma_{H=0}^{N-L}\Sigma_{a_{1}..a_{L}=0}^{p}\Sigma_{j_{1}..j_{H}=p+1}^{M}
       (X^{j_{1}}..X^{j_{H-1}})^{2}\gamma^{a_{L-1}}\langle \partial_{a_{1}}..\partial_{a_{L-1}}\psi^{i},\partial_{a_{1}}..
       \partial_{a_{L}}\psi^{i}\rangle -\nonumber \\
      &&  i\Sigma_{L=0}^{N} \Sigma_{H=0}^{N-L}\Sigma_{a_{1}..a_{L}=0}^{p}\Sigma_{j_{1}..j_{H}=p+1}^{M}\frac{\lambda^{2}}{1.2...N}
      \gamma^{a_{L-1}}(X^{j_{1}}..X^{j_{H}})^{2}\langle \bar{F}^{a_{1}..a_{L-1}}, \bar{F}^{a_{1}..a_{L}}\rangle \Big) \label{P10}
       \end{eqnarray}

   This action is reduced to supersymmetric actions for Mp-brane by puting $N=3$ and $M=10$ in $11$-dimensions
   \cite{k7,k8,k9,k10,k11,k12,k13,k14,k15,k16,k17,k18,k19,k20}. It is clear that number of degrees of freedom for
   bosons and fermions are the same and all particles and their superpartners appear in this action. Also, the
   exact wave equations for scalars, Dirac fields, gauge fields and higher dimensional spinors can be observed in this action.

   Now, we like to extract gravity from actions in BLNA-theory. For this reason, we assume that $A^{ab}$ plays the
   role of graviton and $\bar{A}^{ab}$ has the role of gravitino and other higher dimensional fields have the following
   relations with these fields:

     \begin{eqnarray}
     && A_{a'b'}\rightarrow g_{a'b'} \nonumber \\
      && F_{a'b'c'}=\partial_{[a'} g_{b'c']}=\partial_{a'} g_{b'c'}-\partial_{c'} g_{a'b'}+\partial_{b'} g_{c'a'}\rightarrow
      \Gamma _{a'b'c'}\label{t28}
      \end{eqnarray}

      \begin{eqnarray}
    && A_{a'b'c'}\rightarrow F_{a'b'c'}\rightarrow\Gamma_{a'b'c'} \nonumber \\
      && \tilde{R}_{ca'b'c'}\approx \partial_{[c} A_{a'b'c']} + \langle F_{\lambda ca'},F_{ b'c'}^{\lambda}\rangle
      \approx \partial_{[c}\Gamma _{a'b'c']}+ \Gamma_{\lambda ca'}\Gamma_{ b'c'}^{\lambda}-\Gamma_{\lambda cb'}
      \Gamma_{ a'c'}^{\lambda}\label{t29}
            \end{eqnarray}

      \begin{eqnarray}
                && A_{a'b'c'c}\rightarrow F_{a'b'c'c}\rightarrow \tilde{R}_{a'b'c'c} \nonumber \\
              &&  F_{a_{1},a_{2},..a_{n}}=\partial_{[a_{5}}.. \partial_{a_{n}}\tilde{R}_{a_{1},a_{2},a_{3},a_{4}]}\label{t30}
                \end{eqnarray}

         \begin{eqnarray}
         && \bar{A}_{a'b'}\rightarrow \bar{g}_{a'b'} \nonumber \\
          && \bar{F}_{a'b'c'}=\partial_{[a'} g_{b'c']}=\partial_{a'} g_{b'c'}-\partial_{c'} g_{a'b'}+\partial_{b'} g_{c'a'}
          \rightarrow  \Gamma _{a'b'c'}\label{tt28}
         \end{eqnarray}

          \begin{eqnarray}
            && \bar{A}_{a'b'c'}\rightarrow \bar{F}_{a'b'c'}\rightarrow\Gamma_{a'b'c'} \nonumber \\
     && \bar{R}_{ca'b'c'}\approx \partial_{[c} A_{a'b'c']} + \langle \bar{F}_{\lambda ca'},\bar{F}_{ b'c'}^{\lambda}\rangle
     \approx \partial_{[c}\Gamma _{a'b'c']}+ \Gamma_{\lambda ca'}\Gamma_{ b'c'}^{\lambda}-\Gamma_{\lambda cb'}
     \Gamma_{ a'c'}^{\lambda}\label{tt29}
        \end{eqnarray}

          \begin{eqnarray}
         && \bar{A}_{a'b'c'c}\rightarrow \bar{F}_{a'b'c'c}\rightarrow \bar{R}_{a'b'c'c} \nonumber \\
          &&  \bar{F}_{a_{1},a_{2},..a_{n}}=\partial_{[a_{5}}.. \partial_{a_{n}}\bar{R}_{a_{1},a_{2},a_{3},a_{4}]},
          \label{tt30}
         \end{eqnarray}

  where ($g_{a'b'}$ and $\bar{g}_{a'b'}$) are graviton and gravitino respectively. Substituting equations of
  (\ref{t28},\ref{t29},\ref{t30},\ref{tt28},\ref{tt29},\ref{tt30}) in action of (\ref{P10}), we obtain the action
  of $Gp$-brane in terms of curvatures and metrics:

  \begin{eqnarray}
    &&S_{Gp} = -(T_{Gp}) \int dt \sum_{n=1}^{p}\beta_{n}\Big(
         \delta^{a_{1},a_{2}...a_{n}}_{b_{1}b_{2}....b_{n}}L^{b_{1}}_{a_{1}}...L^{b_{n}}_{a_{n}}\Big)^{1/2}\nonumber\\&&
         (L)^{a_{n}}_{b_{n}}= \delta^{a_{n}}_{b_{n}}Tr\Big(
            \frac{1}{2}\Sigma_{L=0}^{N} \Sigma_{H=0}^{N-L}\Sigma_{a_{1}..a_{L}=0}^{p}\Sigma_{j_{1}..j_{H}=p+1}^{M}
    (X^{j_{1}}..X^{j_{H-1}})^{2}\langle \partial_{a_{1}}..\partial_{a_{L}}X^{i},\partial_{a_{1}}..\partial_{a_{L}}X^{i}\rangle
    + \nonumber \\ &&
     \Sigma_{L=0}^{N} \Sigma_{H=0}^{N-L}\Sigma_{a_{1}..a_{L}=0}^{p}\Sigma_{j_{1}..j_{H}=p+1}^{M}
     \frac{\lambda^{2}}{1.2...N}(X^{j_{1}}..X^{j_{H}})^{2}\langle \partial_{a_{5}}..\partial_{a_{L}}
     \tilde{R}^{a_{1},a_{2},a_{3},a_{4}},\partial_{a_{5}}..\partial_{a_{L}} \tilde{R}^{a_{1},a_{2},a_{3},a_{4}}\rangle  -\nonumber\\ &&
  \frac{1}{2}i\Sigma_{L=0}^{N} \Sigma_{H=0}^{N-L}\Sigma_{a_{1}..a_{L}=0}^{p}\Sigma_{j_{1}..j_{H}=p+1}^{M}(X^{j_{1}}..X^{j_{H-1}})^{2}
  \gamma^{a_{L-1}}\langle \partial_{a_{1}}..\partial_{a_{L-1}}\psi^{i},\partial_{a_{1}}..\partial_{a_{L}}\psi^{i}\rangle -\nonumber \\
    &&  i\Sigma_{L=0}^{N} \Sigma_{H=0}^{N-L}\Sigma_{a_{1}..a_{L}=0}^{p}\Sigma_{j_{1}..j_{H}=p+1}^{M}
    \frac{\lambda^{2}}{1.2...N}\gamma^{a_{L-1}}(X^{j_{1}}..X^{j_{H}})^{2} \nonumber\\ &&
    \langle \partial_{a_{5}}..\partial_{a_{L-1}}
    \bar{R}^{a_{1},a_{2},a_{3},a_{4}},\partial_{a_{5}}..\partial_{a_{L}} \bar{R}^{a_{1},a_{2},a_{3},a_{4}}\rangle +...\Big)
    \label{PO10}
    \end{eqnarray}

  This action contains various orders of curvatures and its derivatives and is reduced to related actions in $F(R)$-gravity
  in \cite{k21,k22} in four dimensional universe. In addition to this, gravity includes both curvatures of gravitons and gravitinoes
  and shows the role of spinor gravity in evolutions of branes. On the other hand, curvatures which are produced by gravitinoes
  have opposite sign respect to curvatures that are created by gravitons; and thus, they may cancel their effects and universe
  seems to be flat. \\

  Now, two significant questions emerge: what is difference between branes and anti-branes physically ? And how are they produced? To
  respond to these questions, we assume that there is nothing at the beginning. Then, two energies with opposite sign are
  emerged such as the sum over them is zero again. After that, these energies  produce $2M$ degrees of freedom which each two
  of them leads to creation of new dimension. At the fourth stage, $M-N$ of degrees of freedom are removed by compacting half
  of $M$-$N$ dimensions on a circle to produce Lie-$N$-algebra. During this compactification, the behaviour of one dimension is
  different with other dimensions for one of initial energies which is known as time. Also, for second energy the behaviour
  of two dimensions is different which leads to the appearance of two time coordinates. After compactification, usual action
  of branes emerge with one time coordinates, however the physics of anti-branes is different and they have two
  times coordinates. \\

   First, we show that two oscillating energies are produced from nothing and expanded in $M^{th}$ dimension. We write:

   \begin{eqnarray}
    && E \equiv 0\equiv E_{1}+E_{2}\equiv 0\equiv N_{1}+N_{2}\equiv k((X^{M})^{2}-  (X^{M})^{2}) = k \int d^{2}x
    (\frac{\partial}{\partial x})^{2}((X^{M})^{2}-  (X^{M})^{2}),
         \label{t1}
         \end{eqnarray}

  where $N_{1/2}$ denote the number of degrees of freedom for first and second energies. These energies oscillate,
  excite and create $M$ dimensions with $2M$ degrees of freedom. We can show this by rewriting Eq. (\ref{t1})as follows:

     \begin{eqnarray}
      && E \equiv 0\equiv k \int d^{2M}x \varepsilon^{i_{1}i_{2}...i_{M}}\varepsilon^{i_{1}i_{2}...i_{M}} (\frac{\partial}
      {\partial x_{i_{1}}}\frac{\partial}{\partial x_{i_{2}}}..\frac{\partial}{\partial x_{i_{M-1}}})^{2}(X^{M})^{2}-
      \nonumber \\ &&  k \int d^{2M}x \varepsilon^{i_{1}i_{2}...i_{M}}\varepsilon^{i_{1}i_{2}...i_{M}} (\frac{\partial}
      {\partial x_{i_{1}}}\frac{\partial}{\partial x_{i_{2}}}..\frac{\partial}{\partial x_{i_{M-1}}})^{2}(X^{M})^{2},
           \label{t2}
           \end{eqnarray}

 where, we use of  $\varepsilon^{i_{1}i_{2}...i_{M}}\varepsilon^{i_{1}i_{2}...i_{M}}=-1$. In Eq. (\ref{t2}), each
 integral and derivative $\int dx \frac{\partial}{\partial x}$ shows one degree of freedom and thus we have $M$ dimensions
 and $2m$ degrees of freedom. Also, each derivative with respect to special dimension of initial energy, produces a new force
 ($F=\frac{\partial V}{\partial x}$) and leads to expansion of energy and creation of new degrees of freedom in that direction.
 We can replace derivatives by brackets as follows \cite{k7,k8,k9,k10,k11,k12,k13,k14,k15,k16,k17,k18,k19,k20}:

    \begin{eqnarray}
    && \frac{\partial}{\partial x_{i_{1}}}X^{M}=[ X^{i_{1}},X^{14}] \nonumber \\ &&
    \frac{\partial}{\partial x_{i_{1}}}\frac{\partial}{\partial x_{i_{2}}}X^{M}=[ X^{i_{1}},X^{i_{2}},X^{M}]\nonumber \\ &&
    (\frac{\partial}{\partial x_{i_{1}}}\frac{\partial}{\partial x_{i_{2}}}..\frac{\partial}{\partial x_{i_{M-1}}})(X^{M})=
    [ X^{i_{1}},X^{i_{2}},...,X^{i_{M-1}},X^{M}]\nonumber \\ &&\varepsilon^{i_{1}i_{2}...i_{M}}\varepsilon^{i_{1}i_{2}...i_{M}}
    (\frac{\partial}{\partial x_{i_{1}}}\frac{\partial}{\partial x_{i_{2}}}..\frac{\partial}{\partial x_{i_{M-1}}})^{2}(X^{M})^{2}=
    \nonumber \\ && \varepsilon^{i_{1}i_{2}...i_{M}}\varepsilon^{i'_{1}i'_{2}...i'_{M}} [(\frac{\partial}{\partial x_{i_{1}}}
    \frac{\partial}{\partial x_{i_{2}}}..\frac{\partial}{\partial x_{i_{M-1}}})(X^{M})][(\frac{\partial}{\partial x_{i'_{1}}}
    \frac{\partial}{\partial x_{i'_{2}}}..\frac{\partial}{\partial x_{i'_{M-1}}})(X^{M})]=\nonumber \\ &&
    \langle [ X_{i_{1}},X_{i_{2}},...,X_{i_{M}}],[ X_{i_{1}},X_{i_{2}},
        ...,X_{i_{M}}] \rangle
           \label{t3}
    \end{eqnarray}

    Using  the mappings of Eq. (\ref{t3}) in Eq. (\ref{t2}), we obtain:

   \begin{eqnarray}
   && E \equiv 0\equiv E_{1}+E_{2}\equiv \nonumber \\ &&  E_{1}=  k \int d^{2M}x \langle [ X_{i_{1}},X_{i_{2}},...,X_{i_{M}}],
   [ X_{i_{1}},X_{i_{2}},...,X_{i_{M}}] \rangle \nonumber \\ &&
   E_{2}=-k \int d^{2M}x \langle [ X_{i_{1}},X_{i_{2}},...,X_{i_{M}}],[ X_{i_{1}},X_{i_{2}},...,X_{i_{M}}] \rangle
           \label{t4}
  \end{eqnarray}

 These energies are similar to action of $Gp$-branes, however it is expected that algebra be of order $N$, however they are
 of order of $M$ and for this reason, we should remove $M-N$ of degrees of freedom by compacting. To this end, we use of the
 mechanism in \cite{k17,k18,k19,k20} and replace $X_{i_{n=1,3,5..M-N}}=i T^{i_{n}}\frac{R}{l_{P}^{1/2}}$   where $l_{P}$
 is the Planck length. We obtain the following action for first energy:

   \begin{eqnarray}
   && E_{1}\equiv k \int d^{2M}x \langle [ X_{i_{1}},X_{i_{2}},...,X_{i_{M}}],
      [X_{i_{1}},X_{i_{2}},...,X_{i_{M}}] \rangle= \nonumber\\&&
   k \int d^{2M}x  \varepsilon^{i_{1}i_{2}...i_{M}}\varepsilon^{i'_{1}i'_{2}...i'_{M}}X_{i_{1}}X_{i_{2}}...X_{i_{M}}X_{i'_{1}}
   X_{i'_{2}}...X_{i'_{M}} = \nonumber \\ && (i)^{2(M-N)}k \int d^{N}x (\frac{R^{M-N}}{l_{P}^{(M-N)/2}})
   \varepsilon^{j_{1}...j_{N}}\varepsilon^{j'_{1}...j'_{N}}X_{j_{1}}...X_{j_{N}}X_{j'_{1}}...X_{j'_{N}}=\nonumber\\ &&
   (i)^{2(M-N)}k \int d^{N}x (\frac{R^{M-N}}{l_{P}^{(M-N)/2}})\langle [ X_{j_{1}},X_{j_{2}},...,X_{j_{N}}],
   [ X_{j_{1}},X_{j_{2}},...,X_{j_{N}}]\rangle=\nonumber\\ && k \int d^{N}x (\frac{R^{M-N}}{l_{P}^{(M-N)/2}})
   \langle [i X_{j_{1}},iX_{j_{2}},...,iX_{j_{M-N}}..,X_{j_{N}}],[ iX_{j_{1}},X_{j_{2}},..,...,iX_{j_{M-N}}.,X_{j_{N}}]\rangle,
            \label{t5}
   \end{eqnarray}

  where we have used $ \varepsilon^{1i_{2}...i_{M}}\varepsilon^{1i'_{2}...i'_{M}}=(-i)^{N-M}\varepsilon^{j_{1}...j_{N}}
  \varepsilon^{j'_{1}...j'_{N}}$. Scalars which are different from other scalars by one extra (i), are located in time directions.
  Thus, in BLNA-theory, we can have $M-N$ time coordinates where $M$ is dimension of world and $N$ is dimension of algebra. For
  an observer on the brane, we can put ($M=p$) where, $p$ is dimension of brane. For example, in $M$-theory, for a four dimensional
  brane like our universe, we have $4$ dimensions and $3$ dimensional algebra. Thus, we  observe only one time coordinates, however
  for branes with higher dimensions, we observe more time coordinates. Also, this equation shows that $N$ should be equal or
  less of ($\frac{M}{2}$) which is consistent with Lie-two algebra in string theory and Lie-three-algebra in $M$-theory. For
  second energy which is different from first one in its sign, we have one extra time coordinate, because we have:

  \begin{eqnarray}
   && E_{2}=-E_{1}=(i)^{2}E_{1}= \nonumber\\ && k \int d^{N}x (\frac{R^{M-N}}{l_{P}^{(M-N)/2}})
   \langle [i X_{j_{1}},iX_{j_{2}},...,iX_{j_{M-N+1}}..,X_{j_{N}}],[ iX_{j_{1}},X_{j_{2}},..,...,iX_{j_{M-N+1}}.,X_{j_{N}}]\rangle
   \label{t6}
  \end{eqnarray}

   Thus, physics of branes which is produced by this energy is different and we have more difference between dimensions.
   For example, in four dimensional universe, we should have two time coordinates and all things are changed. For example,
   in our universe, length of one object can be obtained by $l^{2}=-t^{2}+x_{1}^{2}+x_{2}^{2}+x_{3}^{2}$ where t is time
   and $x_{1}$ are coordinates of space. However in anti-universe, length is defined by
   $\tilde{l}^{2}=-t_{1}^{2}-x_{1}^{2}+x_{2}^{2}+x_{3}^{2}$. Also, energy and momentums which have the relation with
   mass ($m^{2}=-E^{2}+P_{1}^{2}+P_{2}^{2}+P_{3}^{2}$), now their relation is different for anti-universe
   ($m^{2}=-E^{2}-P_{1}^{2}+P_{2}^{2}+P_{3}^{2}$).

    We can correct action in Eq. (\ref{P8}) by regarding ($i^{2(p-N)}$) for branes and ($i^{2(p-N+1)}$) for anti-branes
    and assuming $R=l_{P}^{(1)/2}$. We obtain:

     \begin{eqnarray}
       &&S_{Gp} = -(T_{G0})^{p} \int dt \sum_{n=1}^{p}\beta_{n}\Big(\delta^{a_{1},a_{2}...a_{n}}_{b_{1}b_{2}....b_{n}}
       L^{b_{1}}_{a_{1}}...L^{b_{n}}_{a_{n}}\Big)^{1/2}\nonumber\\&&
           (L)^{a_{n}}_{b_{n}}= \delta^{a_{n}}_{b_{n}}Tr\Big( \Sigma_{L=0}^{N} \Sigma_{H=0}^{N-L}\Sigma_{a_{1}..a_{L}=0}^{p}
           \Sigma_{j_{1}..j_{H}=p+1}^{M}(\nonumber \\
          && i^{2(p-N)}\langle[X^{j_{1}},..X^{j_{H-1}},X^{a_{1}},..X^{a_{L}},X^{j_{H}}],\langle[X^{j_{1}},..X^{j_{H-1}},
          X^{a_{1}},..X^{a_{L}},X^{j_{H}}]\rangle) + \nonumber \\
          && i^{2(p-N)}\Sigma_{L=0}^{N} \Sigma_{H=0}^{N-L}\Sigma_{a_{1}..a_{L}=0}^{p}\Sigma_{j_{1}..j_{H}=p+1}^{M}
         i^{2(p-N)} (\langle[X^{j_{1}},..X^{j_{H}},X^{a_{1}},..X^{a_{L}}],[X^{j_{1}},..X^{j_{H}},X^{a_{1}},..X^{a_{L}}]\rangle)-
         \nonumber \\ &&  i^{2(p-N)+1}\langle[\gamma^{j_{1}},..X^{j_{H-1}},X^{a_{1}},..X^{a_{L}},\psi^{R,j_{H}}],
         \langle[X^{j_{1}},..X^{j_{H-1}},X^{a_{1}},..X^{a_{L}},\psi^{R,j_{H}}]\rangle) - \nonumber \\
           && i^{2(p-N)+1}\Sigma_{L=0}^{N} \Sigma_{H=0}^{N-L}\Sigma_{a_{1}..a_{L}=0}^{p}\Sigma_{j_{1}..j_{H}=p+1}^{M}
         (\langle[\gamma^{j_{1}},..\psi^{R,j_{H}},X^{a_{1}},..X^{a_{L}}],[X^{j_{1}},..\psi^{R,j_{H}},X^{a_{1}},..X^{a_{L}}]
         \rangle)\Big) \label{P88}
           \end{eqnarray}

      \begin{eqnarray}
       &&S_{Anti-Gp} = -(T_{G0})^{p} \int dt \sum_{n=1}^{p}\beta_{n}\Big(\delta^{a_{1},a_{2}...a_{n}}_{b_{1}b_{2}....b_{n}}
      L^{b_{1}}_{a_{1}}...L^{b_{n}}_{a_{n}}\Big)^{1/2}\nonumber\\&&
       (L)^{a_{n}}_{b_{n}}= \delta^{a_{n}}_{b_{n}}Tr\Big( \Sigma_{L=0}^{N} \Sigma_{H=0}^{N-L}\Sigma_{a_{1}..a_{L}=0}^{p}\Sigma_{j_{1}..j_{H}=p+1}^{M}(\nonumber \\
        && i^{2(p-N+1)}\langle[X^{j_{1}},..X^{j_{H-1}},X^{a_{1}},..X^{a_{L}},X^{j_{H}}],\langle[X^{j_{1}},..X^{j_{H-1}},X^{a_{1}},..X^{a_{L}},X^{j_{H}}]\rangle)
        + \nonumber \\  && i^{2(p-N+1)}\Sigma_{L=0}^{N} \Sigma_{H=0}^{N-L}\Sigma_{a_{1}..a_{L}=0}^{p}\Sigma_{j_{1}..j_{H}=p+1}^{M}
             i^{2(p-N)} (\langle[X^{j_{1}},..X^{j_{H}},X^{a_{1}},..X^{a_{L}}],[X^{j_{1}},..X^{j_{H}},X^{a_{1}},..X^{a_{L}}]\rangle)- \nonumber \\ &&
   i^{2(p-N+1)+1}\langle[\gamma^{j_{1}},..X^{j_{H-1}},X^{a_{1}},..X^{a_{L}},\psi^{R,j_{H}}],\langle[X^{j_{1}},..X^{j_{H-1}},X^{a_{1}},..X^{a_{L}},\psi^{R,j_{H}}]\rangle)
   - \nonumber \\
          && i^{2(p-N+1)+1}\Sigma_{L=0}^{N} \Sigma_{H=0}^{N-L}\Sigma_{a_{1}..a_{L}=0}^{p}\Sigma_{j_{1}..j_{H}=p+1}^{M}
            (\langle[\gamma^{j_{1}},..\psi^{R,j_{H}},X^{a_{1}},..X^{a_{L}}],[X^{j_{1}},..\psi^{R,j_{H}},X^{a_{1}},..X^{a_{L}}]
            \rangle)\Big) \label{PP8}
    \end{eqnarray}

  Using the laws in Eq. (\ref{P9}) and replacing gauge fields by mappings in  Eqs. (\ref{t28}-\ref{tt30})
  % (\ref{t28},\ref{t29},\ref{t30},\ref{tt28},\ref{tt29},\ref{tt30}),
  we can rewrite action of (\ref{PO10}) and obtain the action of $Gp$-brane and anti-$Gp$-brane
  in terms of curvatures:

    \begin{eqnarray}
    &&S_{Gp} = -(T_{Gp}) \int dt \sum_{n=1}^{p}\beta_{n}\Big(\delta^{a_{1},a_{2}...a_{n}}_{b_{1}b_{2}....b_{n}}L^{b_{1}}_{a_{1}}...
    L^{b_{n}}_{a_{n}}\Big)^{1/2}\nonumber \\ &&
    (L)^{a_{n}}_{b_{n}}= \delta^{a_{n}}_{b_{n}}Tr\Big(\frac{1}{2}i^{2(p-N)}\Sigma_{L=0}^{N} \Sigma_{H=0}^{N-L}\Sigma_{a_{1}..
    a_{L}=0}^{p} Sigma_{j_{1}..j_{H}=p+1}^{M}(X^{j_{1}}..X^{j_{H-1}})^{2}\langle \partial_{a_{1}}..\partial_{a_{L}}X^{i},
    \partial_{a_{1}}..\partial_{a_{L}}X^{i}\rangle
          + \nonumber \\ &&
      i^{2(p-N)} \Sigma_{L=0}^{N} \Sigma_{H=0}^{N-L}\Sigma_{a_{1}..a_{L}=0}^{p}\Sigma_{j_{1}..j_{H}=p+1}^{M}
      \frac{\lambda^{2}}{1.2...N}(X^{j_{1}}..X^{j_{H}})^{2}
      \langle \partial_{a_{5}}..\partial_{a_{L}}\tilde{R}^{a_{1},a_{2},a_{3},a_{4}},\partial_{a_{5}}..\partial_{a_{L}}
      \tilde{R}^{a_{1},a_{2},a_{3},a_{4}}\rangle
      - \nonumber\\ &&
      \frac{1}{2}i^{2(p-N)+1}\Sigma_{L=0}^{N} \Sigma_{H=0}^{N-L}\Sigma_{a_{1}..a_{L}=0}^{p}\Sigma_{j_{1}..j_{H}=p+1}^{M}
      (X^{j_{1}}..X^{j_{H-1}})^{2}
      \gamma^{a_{L-1}}\langle \partial_{a_{1}}..\partial_{a_{L-1}}\psi^{i},\partial_{a_{1}}..\partial_{a_{L}}\psi^{i}\rangle -
      \nonumber \\
  &&  i^{2(p-N)+1}\Sigma_{L=0}^{N} \Sigma_{H=0}^{N-L}\Sigma_{a_{1}..a_{L}=0}^{p}\Sigma_{j_{1}..j_{H}=p+1}^{M}
  \frac{\lambda^{2}}{1.2...N}
      \gamma^{a_{L-1}}(X^{j_{1}}..X^{j_{H}})^{2} \nonumber \\ &&
      \langle \partial_{a_{5}}..\partial_{a_{L-1}}\bar{R}^{a_{1},a_{2},a_{3},a_{4}},
      \partial_{a_{5}}..\partial_{a_{L}}
      \bar{R}^{a_{1},a_{2},a_{3},a_{4}}\rangle +...\Big) \label{PPP10}
   \end{eqnarray}

     \begin{eqnarray}
     && S_{Anti-Gp} = -(T_{Anti-Gp}) \int dt \sum_{n=1}^{p}\beta_{n}\Big(
       \delta^{a_{1},a_{2}...a_{n}}_{b_{1}b_{2}....b_{n}}L^{b_{1}}_{a_{1}}...L^{b_{n}}_{a_{n}}\Big)^{1/2}\nonumber\\&&
     (L)^{a_{n}}_{b_{n}}= \delta^{a_{n}}_{b_{n}}Tr\Big(\frac{1}{2}i^{2(p-N+1)}\Sigma_{L=0}^{N} \Sigma_{H=0}^{N-L}
     \Sigma_{a_{1}..a_{L}=0}^{p}
          \Sigma_{j_{1}..j_{H}=p+1}^{M}(X^{j_{1}}..X^{j_{H-1}})^{2}\langle \partial_{a_{1}}..\partial_{a_{L}}X^{i},
          \partial_{a_{1}}..\partial_{a_{L}}X^{i}\rangle
          + \nonumber \\  &&
       i^{2(p-N+1)} \Sigma_{L=0}^{N} \Sigma_{H=0}^{N-L}\Sigma_{a_{1}..a_{L}=0}^{p}\Sigma_{j_{1}..j_{H}=p+1}^{M}
       \frac{\lambda^{2}}{1.2...N}(X^{j_{1}}..X^{j_{H}})^{2}
       \langle \partial_{a_{5}}..\partial_{a_{L}}\hat{\tilde{R}}^{a_{1},a_{2},a_{3},a_{4}},\partial_{a_{5}}..\partial_{a_{L}}
       \hat{\tilde{R}}^{a_{1},a_{2},a_{3},a_{4}}\rangle  -\nonumber\\ &&
       \frac{1}{2}i^{2(p-N+1)+1}\Sigma_{L=0}^{N} \Sigma_{H=0}^{N-L}\Sigma_{a_{1}..a_{L}=0}^{p}\Sigma_{j_{1}..j_{H}=p+1}^{M}
       (X^{j_{1}}..X^{j_{H-1}})^{2}
       \gamma^{a_{L-1}}\langle \partial_{a_{1}}..\partial_{a_{L-1}}\psi^{i},\partial_{a_{1}}..\partial_{a_{L}}
       \psi^{i}\rangle -\nonumber \\
        &&  i^{2(p-N+1)+1}\Sigma_{L=0}^{N} \Sigma_{H=0}^{N-L}\Sigma_{a_{1}..a_{L}=0}^{p}\Sigma_{j_{1}..j_{H}=p+1}^{M}
        \frac{\lambda^{2}}{1.2...N}
        \gamma^{a_{L-1}}(X^{j_{1}}..X^{j_{H}})^{2}\nonumber \\ &&
        \langle \partial_{a_{5}}..\partial_{a_{L-1}}\hat{\bar{R}}^{a_{1},a_{2},a_{3},a_{4}},
        \partial_{a_{5}}..\partial_{a_{L}} \hat{\bar{R}}^{a_{1},a_{2},a_{3},a_{4}}\rangle +...\Big), \label{PP10}
      \end{eqnarray}

  where $\hat{\tilde{R}}$ and $\hat{\bar{R}}$  are curvatures of graviton and gravitino in anti-branes respectively.
  These curvatures are different from curvatures in branes, because they contain more derivatives respect to time coordinates.
  In addition to this, the sign of curvatures in branes are reversed with respect to anti-branes, which means that the lines of gravity go
  outside the branes, while these lines go inside the anti-branes and thus these objects attract each other. For four dimensional
  universe in 11 dimensional $M$-theory with Lie-three-algebra the action is reduced to known
  actions for $F(R)$-gravity within \cite{k21,k22}. \\

  Now, we will show that gravitons and gravitinoes produce two different wormholes that act reverse to each other. The wormhole
  which is produced by gravitons prevents the getting away of quarks and anti-quarks from each other and generates confinement,
  while the wormhole, which is produced by gravitinoes, prevents quarks and anti-quarks from coming close to each other and creates
  deconfinement. For this reason, in quarkonium, quarks and anti-quarks don't go away from each other or come close to each other.
  To obtain the shape of wormholes, we use the method in \cite{k23} and obtain the momentum density. However before doing
  it, we define bosonic and fermionic Lagrangian as:

    \begin{eqnarray}
     &&S_{Gp} = -(T_{Gp}) \int dt \sum_{n=1}^{p}\beta_{n}
     \Big(\delta^{a_{1},a_{2}...a_{n}}_{b_{1}b_{2}....b_{n}}L^{b_{1}}_{a_{1}}...L^{b_{n}}_{a_{n}}\Big)^{1/2}
     \nonumber\\&&
     %\nonumber\\&&\nonumber\\&&
     (L)^{a_{n}}_{b_{n},brane}=(L)^{a_{n}}_{b_{n},bosonic,brane}+(L)^{a_{n}}_{b_{n},fermionic,brane}
     \nonumber\\&&
     \nonumber\\&&\nonumber\\&&
      (L)^{a_{n}}_{b_{n},bosonic,brane}= \delta^{a_{n}}_{b_{n}}Tr
      \Big(\frac{1}{2}i^{2(p-N)}\Sigma_{L=0}^{N} \Sigma_{H=0}^{N-L}\Sigma_{a_{1}..a_{L}=0}^{p}
      \Sigma_{j_{1}..j_{H}=p+1}^{M}(X^{j_{1}}..X^{j_{H-1}})^{2}
      \langle \partial_{a_{1}}..\partial_{a_{L}}X^{i},\partial_{a_{1}}..\partial_{a_{L}}X^{i}\rangle + \nonumber \\ &&
       i^{2(p-N)} \Sigma_{L=0}^{N} \Sigma_{H=0}^{N-L}\Sigma_{a_{1}..a_{L}=0}^{p}\Sigma_{j_{1}..j_{H}=p+1}^{M}
       \frac{\lambda^{2}}{1.2...N}(X^{j_{1}}..X^{j_{H}})^{2}\langle \partial_{a_{5}}..\partial_{a_{L}}
       \tilde{R}^{a_{1},a_{2},a_{3},a_{4}},\partial_{a_{5}}..\partial_{a_{L}} \tilde{R}^{a_{1},a_{2},a_{3},a_{4}}\rangle+... \Big)
       \nonumber\\ &&
       \nonumber \\ &&\nonumber \\ &&
       (L)^{a_{n}}_{b_{n},fermionic,brane}= \delta^{a_{n}}_{b_{n}}Tr
       \Big(-\frac{1}{2}i^{2(p-N)+1}\Sigma_{L=0}^{N} \Sigma_{H=0}^{N-L}\Sigma_{a_{1}..a_{L}=0}^{p}\Sigma_{j_{1}..j_{H}=p+1}^{M}
       \times \nonumber \\ &&
       (X^{j_{1}}..X^{j_{H-1}})^{2}\gamma^{a_{L-1}}\langle \partial_{a_{1}}..\partial_{a_{L-1}}\psi^{i},\partial_{a_{1}}..
       \partial_{a_{L}}\psi^{i}\rangle -\nonumber \\ &&
       i^{2(p-N)+1}\Sigma_{L=0}^{N} \Sigma_{H=0}^{N-L}\Sigma_{a_{1}..a_{L}=0}^{p}\Sigma_{j_{1}..j_{H}=p+1}^{M}
       \frac{\lambda^{2}}{1.2...N}\gamma^{a_{L-1}}(X^{j_{1}}..X^{j_{H}})^{2}
       \nonumber \\ &&
       \langle \partial_{a_{5}}..\partial_{a_{L-1}}\bar{R}^{a_{1},a_{2},a_{3},a_{4}},\partial_{a_{5}}..\partial_{a_{L}}
       \bar{R}^{a_{1},a_{2},a_{3},a_{4}}\rangle +...\Big) \label{kp1}
     \end{eqnarray}
   and
  \begin{eqnarray}
       &&S_{Anti-Gp} = -(T_{Anti-Gp}) \int dt \sum_{n=1}^{p}\beta_{n}\Big(
      \delta^{a_{1},a_{2}...a_{n}}_{b_{1}b_{2}....b_{n}}L^{b_{1}}_{a_{1}}...L^{b_{n}}_{a_{n}}\Big)^{1/2}
      \nonumber\\&&
      \nonumber\\&&\nonumber\\&&
      (L)^{a_{n}}_{b_{n},anti-brane}=(L)^{a_{n}}_{b_{n},bosonic,anti-brane}+(L)^{a_{n}}_{b_{n},fermionic,anti-brane}
      \nonumber\\&&
      \nonumber\\&&\nonumber\\&&
       (L)^{a_{n}}_{b_{n},bosonic,anti-brane}= \delta^{a_{n}}_{b_{n}}Tr\Big(
    \frac{1}{2}i^{2(p-N+1)}\Sigma_{L=0}^{N} \Sigma_{H=0}^{N-L}\Sigma_{a_{1}..a_{L}=0}^{p}\Sigma_{j_{1}..j_{H}=
    p+1}^{M}(X^{j_{1}}..X^{j_{H-1}})^{2}\langle \partial_{a_{1}}..\partial_{a_{L}}X^{i},\partial_{a_{1}}..\partial_{a_{L}}X^{i}\rangle
    + \nonumber \\  &&
    i^{2(p-N)} \Sigma_{L=0}^{N} \Sigma_{H=0}^{N-L}\Sigma_{a_{1}..a_{L}=0}^{p}\Sigma_{j_{1}..j_{H}=p+1}^{M}
    \frac{\lambda^{2}}{1.2...N}(X^{j_{1}}..X^{j_{H}})^{2}\langle \partial_{a_{5}}..\partial_{a_{L}}
    \hat{\tilde{R}}^{a_{1},a_{2},a_{3},a_{4}},\partial_{a_{5}}..\partial_{a_{L}}
    \hat{\tilde{R}}^{a_{1},a_{2},a_{3},a_{4}}\rangle+... \Big) \nonumber \\ &&
    \nonumber \\ &&\nonumber \\  &&
    (L)^{a_{n}}_{b_{n},fermionic,anti-brane}= \delta^{a_{n}}_{b_{n}}Tr\Big(
    - \frac{1}{2}i^{2(p-N+1)+1}\Sigma_{L=0}^{N} \Sigma_{H=0}^{N-L}\Sigma_{a_{1}..a_{L}=0}^{p}\Sigma_{j_{1}..j_{H}=p+1}^{M}\times
    \nonumber \\ &&
    (X^{j_{1}}..X^{j_{H-1}})^{2}\gamma^{a_{L-1}}\langle \partial_{a_{1}}..\partial_{a_{L-1}}\psi^{i},\partial_{a_{1}}..\partial_{a_{L}}
    \psi^{i}\rangle -\nonumber \\ &&
    i^{2(p-N)+1}\Sigma_{L=0}^{N} \Sigma_{H=0}^{N-L}\Sigma_{a_{1}..a_{L}=0}^{p}\Sigma_{j_{1}..j_{H}=p+1}^{M}\frac{\lambda^{2}}{1.2...N}
    \gamma^{a_{L-1}}(X^{j_{1}}..X^{j_{H}})^{2}
    \nonumber \\ &&
    \langle \partial_{a_{5}}..\partial_{a_{L-1}}
    \hat{\bar{R}}^{a_{1},a_{2},a_{3},a_{4}},\partial_{a_{5}}..\partial_{a_{L}} \hat{\bar{R}}^{a_{1},a_{2},a_{3},a_{4}}\rangle +...\Big)
    \label{kp2}
  \end{eqnarray}

    First, we should calculate the momentum densities for bosonic part and fermionic part of Lagrangian in Eq. (\ref{kp1})
    respect to derivatives of curvature. To this end, we begin with derivatives of order of $p-4$, where $p$ is dimension of
    brane and $4$ denotes four indices of curvature. We obtain

     \begin{eqnarray}
       && \Pi_{bosonic,brane,p-4}\approx \frac{ i^{2(p-N)} \Sigma_{L=0}^{N} \Sigma_{H=0}^{N-L}\Sigma_{a_{1}..a_{L}=0}^{p}
       \Sigma_{j_{1}..j_{H}=p+1}^{M}\frac{\lambda^{2}}{1.2...N}(X^{j_{1}}..X^{j_{H}})^{2}\partial_{a_{5}}..\partial_{a_{L}}
       \tilde{R}^{a_{1},a_{2},a_{3},a_{4}}}{\sqrt{(L)^{a_{n}}_{b_{n},bosonic,brane}}}
       \label{kp3}
       \end{eqnarray}

       \begin{eqnarray}
     && \Pi_{fermionic,brane,p-4}\approx \frac{ -i^{2(p-N)+1} \Sigma_{L=0}^{N} \Sigma_{H=0}^{N-L}\Sigma_{a_{1}..a_{L}=0}^{p}
     \Sigma_{j_{1}..j_{H}=p+1}^{M}\frac{\lambda^{2}}{1.2...N}(X^{j_{1}}..X^{j_{H}})^{2}\partial_{a_{5}}..\partial_{a_{L}}
     \bar{R}^{a_{1},a_{2},a_{3},a_{4}}}{\sqrt{(L)^{a_{n}}_{b_{n},bosonic,brane}}}
     \label{kp4}
       \end{eqnarray}

      \begin{eqnarray}
      && \Pi_{bosonic,anti-brane,p-4}\approx \frac{ i^{2(p-N+1)} \Sigma_{L=0}^{N} \Sigma_{H=0}^{N-L}\Sigma_{a_{1}..a_{L}=0}^{p}
      \Sigma_{j_{1}..j_{H}=p+1}^{M}\frac{\lambda^{2}}{1.2...N}(X^{j_{1}}..X^{j_{H}})^{2}\partial_{a_{5}}..\partial_{a_{L}}
      \bar{\tilde{R}}^{a_{1},a_{2},a_{3},a_{4}}}{\sqrt{(L)^{a_{n}}_{b_{n},bosonic,brane}}} \label{kp5}
      \end{eqnarray}

    \begin{eqnarray}
    && \Pi_{fermionic,anti-brane,p-4}\approx \nonumber \\ &&
    \frac{ -i^{2(p-N+1)+1} \Sigma_{L=0}^{N} \Sigma_{H=0}^{N-L}\Sigma_{a_{1}..a_{L}=0}^{p}
    \Sigma_{j_{1}..j_{H}=p+1}^{M}\frac{\lambda^{2}}{1.2...N}(X^{j_{1}}..X^{j_{H}})^{2}\partial_{a_{5}}..\partial_{a_{L}}
    \hat{\bar{R}}^{a_{1},a_{2},a_{3},a_{4}}}{\sqrt{(L)^{a_{n}}_{b_{n},bosonic,brane}}} \label{kp6}
     \end{eqnarray}

    We assume that all coordinates are the same ($x^{2..p}=\sigma$) and construct a p dimensional sphere. The Hamiltonian
    for this system can be obtained as:

     \begin{eqnarray}
     &&H_{p-4}^{1}\approx 4\pi\int d\sigma \sigma^{p-1}\Sigma_{L=0}^{N} \Sigma_{H=0}^{N-L}\Sigma_{a_{1}..a_{L}=0}^{p}
     \Sigma_{j_{1}..j_{H}=p+1}^{M}\frac{\lambda^{2}}{1.2...N}(X^{j_{1}}..X^{j_{H}})^{2}\partial_{t}..\partial_{a_{L}}
     \tilde{R}^{a_{1},a_{2},a_{3},a_{4}} \Pi_{bosonic,brane,p-4}i^{2(p-N)} \nonumber\\&&
     -4\pi\int d\sigma \sigma^{p-1} \Sigma_{L=0}^{N} \Sigma_{H=0}^{N-L}\Sigma_{a_{1}..a_{L}=0}^{p}\Sigma_{j_{1}..j_{H}=p+1}^{M}
     \frac{\lambda^{2}}{1.2...N}(X^{j_{1}}..X^{j_{H}})^{2}\partial_{t}..\partial_{a_{L}} \bar{R}^{a_{1},a_{2},a_{3},a_{4}}
     \Pi_{fermionic,brane,p-4}i^{2(p-N)+1} \nonumber\\&& +4\pi\int d\sigma \sigma^{p-1}\Sigma_{L=0}^{N} \Sigma_{H=0}^{N-L}
     \Sigma_{a_{1}..a_{L}=0}^{p}\Sigma_{j_{1}..j_{H}=p+1}^{M}\frac{\lambda^{2}}{1.2...N}(X^{j_{1}}..X^{j_{H}})^{2}\partial_{t}..
     \partial_{a_{L}} \hat{\tilde{R}}^{a_{1},a_{2},a_{3},a_{4}} \Pi_{bosonic,anti-brane,p-4}i^{2(p-N+1)}\nonumber\\&&
     -4\pi\int d\sigma \sigma^{p-1} \Sigma_{L=0}^{N} \Sigma_{H=0}^{N-L}\Sigma_{a_{1}..a_{L}=0}^{p}\Sigma_{j_{1}..j_{H}=p+1}^{M}
     \frac{\lambda^{2}}{1.2...N}(X^{j_{1}}..X^{j_{H}})^{2}\partial_{t}..\partial_{a_{L}} \hat{\bar{R}}^{a_{1},a_{2},a_{3},a_{4}}
     \Pi_{fermionic,anti-brane,p-4}i^{2(p-N+1)+1} \nonumber\\&&-L_{bosonic,brane,p-4}^{1}-L_{fermionic,brane,p-4}^{
 1}\nonumber\\&&-L_{bosonic,anti-brane,p-4}^{1}-L_{fermionic,anti-brane,p-4}^{1}=\nonumber\\&&4\pi\int d\sigma \sigma^{p-1}
 \Sigma_{L=0}^{N} \Sigma_{H=0}^{N-L}\Sigma_{a_{1}..a_{L}=0}^{p}\Sigma_{j_{1}..j_{H}=p+1}^{M}\frac{\lambda^{2}}{1.2...N}(X^{j_{1}}..
 X^{j_{H}})^{2}\partial_{a_{5}}..\partial_{a_{L}} \tilde{R}^{a_{1},a_{2},a_{3},a_{4}} \Pi_{bosonic,brane,p-4}i^{2(p-N)}
 \nonumber\\&&
     -4\pi\int d\sigma \sigma^{p-1} \Sigma_{L=0}^{N} \Sigma_{H=0}^{N-L}\Sigma_{a_{1}..a_{L}=0}^{p}\Sigma_{j_{1}..j_{H}=p+1}^{M}
     \frac{\lambda^{2}}{1.2...N}(X^{j_{1}}..X^{j_{H}})^{2}\partial_{a_{5}}..\partial_{a_{L}} \bar{R}^{a_{1},a_{2},a_{3},a_{4}}
     \Pi_{fermionic,brane,p-4}i^{2(p-N)+1} \nonumber\\&& +4\pi\int d\sigma \sigma^{p-1}\Sigma_{L=0}^{N} \Sigma_{H=0}^{N-L}
     \Sigma_{a_{1}..a_{L}=0}^{p}\Sigma_{j_{1}..j_{H}=p+1}^{M}\frac{\lambda^{2}}{1.2...N}(X^{j_{1}}..X^{j_{H}})^{2}
     \partial_{a_{5}}..\partial_{a_{L}} \hat{\tilde{R}}^{a_{1},a_{2},a_{3},a_{4}} \Pi_{bosonic,anti-brane,p-4}i^{2(p-N+1)}
     \nonumber\\&&
     +4\pi\int d\sigma  \Sigma_{L=0}^{N} \Sigma_{H=0}^{N-L}\Sigma_{a_{1}..a_{L}=0}^{p}\Sigma_{j_{1}..j_{H}=p+1}^{M}
     \frac{\lambda^{2}}{1.2...N}(X^{j_{1}}..X^{j_{H}})^{2}\partial_{a_{5}}..\partial_{a_{L}} \hat{\bar{R}}^{a_{1},a_{2},a_{3},a_{4}}
     \Pi_{fermionic,anti-brane,p-4}i^{2(p-N+1)+1} \nonumber\\&&-4\pi\int d\sigma \sigma^{p-1}\Sigma_{L=0}^{N} \Sigma_{H=0}^{N-L}
     \Sigma_{a_{1
 }..a_{L}=0}^{p}\Sigma_{j_{1}..j_{H}=p+1}^{M}\frac{\lambda^{2}}{1.2...N}(X^{j_{1}}..X^{j_{H}})^{2}\partial_{a_{6}}..
 \partial_{a_{L}} \tilde{R}^{a_{1},a_{2},a_{3},a_{4}}\times\nonumber\\&&
 \partial_{a_{5}}(\sigma^{p-1} \Pi_{bosonic,brane,p-4}i^{2(p-N)}) \nonumber\\&&
     -4\pi\int d\sigma  \Sigma_{L=0}^{N} \Sigma_{H=0}^{N-L}\Sigma_{a_{1}..a_{L}=0}^{p}\Sigma_{j_{1}..j_{H}=p+1}^{M}
     \frac{\lambda^{2}}{1.2...N}(X^{j_{1}}..X^{j_{H}})^{2}\partial_{a_{6}}..\partial_{a_{L}} \bar{R}^{a_{1},a_{2},a_{3},a_{4}}
     \times\nonumber\\&&\partial_{a_{5}}(\sigma^{p-1}\Pi_{fermionic,brane,p-4}i^{2(p-N)+1} )\nonumber\\&&
     -4\pi\int d\sigma \Sigma_{L=0}^{N} \Sigma_{H=0}^{N-L}\Sigma_{a_{1}..a_{L}=0}^{p}\Sigma_{j_{1}..j_{H}=p+1}^{M}
     \frac{\lambda^{2}}{1.2...N}(X^{j_{1}}..X^{j_{H}})^{2}\partial_{a_{6}}..\partial_{a_{L}}
     \hat{\tilde{R}}^{a_{1},a_{2},a_{3},a_{4}}\times\nonumber\\&& \partial_{a_{5}}(\sigma^{p-1}
     \Pi_{bosonic,anti-brane,p-4}i^{2(p-N+1)}) \nonumber\\&&+4\pi\int d\sigma  \Sigma_{L=0}^{N} \Sigma_{H=0}^{N-L}
     \Sigma_{a_{1}..a_{L}=0}^{p}\Sigma_{j_{1}..j_{H}=p+1}^{M}\frac{\lambda^{2}}{1.2...N}(X^{j_{1}}..X^{j_{H}})^{2}
     \partial_{a_{6}}..\partial_{a_{L}} \hat{\bar{R}}^{a_{1},a_{2},a_{3},a_{4}}\times\nonumber\\&&\partial_{a_{5}}
 (\sigma^{p-1}\Pi_{fermionic,anti-brane,p-4}i^{2(p-N+1)+1})  \nonumber\\&&-L_{bosonic,brane,p-4}^{1}-L_{fermionic,brane,p-4}^{1}
 \nonumber\\&&-L_{bosonic,anti-brane,p-4}^{1}-L_{fermionic,anti-brane,p-4}^{1},
             \label{kp7}
             \end{eqnarray}

     where we have used in the second step integrated by parts. We can impose the constraint
     ($\partial_{a_{5}}(\sigma^{p-1}\Pi_{bosonic/fermionic,brane/anti-brane,p-4}i^{2(p-N)})/i^{2(p-N+1)+1}) =0$)
     and obtain the momentum densities:

     \begin{eqnarray}
      && \Pi_{bosonic,brane,p-4}=\frac{i^{2(p-N)}k_{bosonic,brane,p-4}}{\sigma^{p-1}}\nonumber\\&&
      \Pi_{fermionic,brane,p-4}=-\frac{i^{2(p-N)+1}k_{fermionic,brane,p-4}}{\sigma^{p-1}}\nonumber\\&&
      \Pi_{bosonic,anti-brane,p-4}=\frac{i^{2(p-N+1)}k_{bosonic,anti-brane,p-4}}{\sigma^{p-1}}\nonumber\\&&
      \Pi_{fermionic,anti-brane,p-4}=-\frac{i^{2(p-N+1)+1}k_{fermionic,anti-brane,p-4}}{\sigma^{p-1}}
      \label{kp8}
      \end{eqnarray}

 Using momentum densities in Eqs. (\ref{kp8}) and (\ref{kp7}), we can calculate the Hamiltonian as:

       \begin{eqnarray}
         &&
        H_{p-4}^{1}\approx 4\pi\int d\sigma \sigma^{p-1}\Big([ \frac{1}{2}i^{2(p-N)}\Sigma_{L=0}^{N} \Sigma_{H=0}^{N-L}
        \Sigma_{a_{1}..a_{L}=0}^{p}\Sigma_{j_{1}..j_{H}=p+1}^{M}(X^{j_{1}}..X^{j_{H-1}})^{2}\langle \partial_{a_{1}}..
        \partial_{a_{L}}X^{i},\partial_{a_{1}}..\partial_{a_{L}}X^{i}\rangle + \nonumber \\
         &&
        i^{2(p-N)} \Sigma_{L=0}^{N} \Sigma_{H=0}^{N-L}\Sigma_{a_{1}..a_{L}=0}^{p}\Sigma_{j_{1}..j_{H}=p+1}^{M}
        \frac{\lambda^{2}}{1.2...N}(X^{j_{1}}..X^{j_{H}})^{2}\langle \partial_{a_{6}}..\partial_{a_{L}}
        \tilde{R}^{a_{1},a_{2},a_{3},a_{4}},\partial_{a_{6}}..\partial_{a_{L}} \tilde{R}^{a_{1},a_{2},a_{3},a_{4}}
        \rangle+... ]^{1/2}\times\nonumber \\
                                                                                                    &&\sqrt{1+(\frac{i^{2(p-N)}k_{bosonic,brane,p-4}}{\sigma^{p-1}})^{2}}+\nonumber \\&&
       [-\frac{1}{2}i^{2(p-N)+1}\Sigma_{L=0}^{N} \Sigma_{H=0}^{N-L}\Sigma_{a_{1}..a_{L}=0}^{p}\Sigma_{j_{1}..j_{H}=p+1}^{M}
       (X^{j_{1}}..X^{j_{H-1}})^{2}\gamma^{a_{L-1}}\langle \partial_{a_{1}}..\partial_{a_{L-1}}\psi^{i},\partial_{a_{1}}..
       \partial_{a_{L}}\psi^{i}\rangle -\nonumber \\
       &&  i^{2(p-N)+1}\Sigma_{L=0}^{N} \Sigma_{H=0}^{N-L}\Sigma_{a_{1}..a_{L}=0}^{p}\Sigma_{j_{1}..j_{H}=p+1}^{M}
        \frac{\lambda^{2}}{1.2...N}\gamma^{a_{L-1}}(X^{j_{1}}..X^{j_{H}})^{2}\langle \partial_{a_{6}}..\partial_{a_{L-1}}
        \bar{R}^{a_{1},a_{2},a_{3},a_{4}},\partial_{a_{6}}..\partial_{a_{L}} \bar{R}^{a_{1},a_{2},a_{3},a_{4}}\rangle +...]^{1/2}
        \times  \nonumber \\&& \sqrt{1+(-\frac{i^{2(p-N)+1}k_{fermionic,brane,p-4}}{\sigma^{p-1}})^{2}}+ \nonumber \\&&
        [ \frac{1}{2}i^{2(p-N+1)}\Sigma_{L=0}^{N} \Sigma_{H=0}^{N-L}\Sigma_{a_{1}..a_{L}=0}^{p}\Sigma_{j_{1}..j_{H}=p+1}^{M}
        (X^{j_{1}}..X^{j_{H-1}})^{2}\langle \partial_{a_{1}}..\partial_{a_{L}}X^{i},\partial_{a_{1}}..\partial_{a_{L}}X^{i}
        \rangle + \nonumber \\
        &&
      i^{2(p-N)} \Sigma_{L=0}^{N} \Sigma_{H=0}^{N-L}\Sigma_{a_{1}..a_{L}=0}^{p}\Sigma_{j_{1}..j_{H}=p+1}^{M}
      \frac{\lambda^{2}}{1.2...N}(X^{j_{1}}..X^{j_{H}})^{2}\langle \partial_{a_{6}}..\partial_{a_{L}}
      \hat{\tilde{R}}^{a_{1},a_{2},a_{3},a_{4}},\partial_{a_{6}}..\partial_{a_{L}} \hat{\tilde{R}}^{a_{1},a_{2},a_{3},a_{4}}
      \rangle+... ]^{1/2}\times\nonumber \\&&\sqrt{1+(\frac{i^{2(p-N+1)}k_{bosonic,anti-brane,p-4}}{\sigma^{p-1}})^{2}} +
      \nonumber \\&& [  - \frac{1}{2}i^{2(p-N+1)+1}
      \Sigma_{L=0}^{N} \Sigma_{H=0}^{N-L}\Sigma_{a_{1}..a_{L}=0}^{p}\Sigma_{j_{1}..j_{H}=p+1}^{M}(X^{j_{1}}..X^{j_{H-1}})^{2}
      \gamma^{a_{L-1}}\langle \partial_{a_{1}}..\partial_{a_{L-1}}\psi^{i},\partial_{a_{1}}..\partial_{a_{L}}\psi^{i}\rangle -
      \nonumber \\
       &&  i^{2(p-N)+1}\Sigma_{L=0}^{N} \Sigma_{H=0}^{N-L}\Sigma_{a_{1}..a_{L}=0}^{p}\Sigma_{j_{1}..j_{H}=p+1}^{M}
       \frac{\lambda^{2}}{1.2...N}\gamma^{a_{L-1}}(X^{j_{1}}..X^{j_{H}})^{2}\langle \partial_{a_{6}}..\partial_{a_{L-1}}
       \hat{\bar{R}}^{a_{1},a_{2},a_{3},a_{4}},\partial_{a_{6}}..\partial_{a_{L}} \hat{\bar{R}}^{a_{1},a_{2},a_{3},a_{4}}
       \rangle +..]^{1/2}\times\nonumber \\
       &&\sqrt{1+(-\frac{i^{2(p-N+1)+1}k_{fermionic,anti-brane,p-4}}{\sigma^{p-1}})^{2}}\Big)\label{kp9}
       \end{eqnarray}

  We use of previous mechanism again and obtain momentum densities for curvature of order $p-5$:

   \begin{eqnarray}
   && \Pi_{bosonic,brane,p-5}\approx \frac{ i^{2(p-N)} \Sigma_{L=0}^{N} \Sigma_{H=0}^{N-L}\Sigma_{a_{1}..a_{L}=0}^{p}
   \Sigma_{j_{1}..j_{H}=p+1}^{M}\frac{\lambda^{2}}{1.2...N}(X^{j_{1}}..X^{j_{H}})^{2}\partial_{a_{6}}..\partial_{a_{L}}
   \tilde{R}^{a_{1},a_{2},a_{3},a_{4}}}{H_{p-4}^{1}} \label{kp10}
   \end{eqnarray}

    \begin{eqnarray}
   && \Pi_{fermionic,brane,p-5}\approx \frac{ -i^{2(p-N)+1} \Sigma_{L=0}^{N} \Sigma_{H=0}^{N-L}\Sigma_{a_{1}..a_{L}=0}^{p}
   \Sigma_{j_{1}..j_{H}=p+1}^{M}\frac{\lambda^{2}}{1.2...N}(X^{j_{1}}..X^{j_{H}})^{2}\partial_{a_{6}}..\partial_{a_{L}}
   \bar{R}^{a_{1},a_{2},a_{3},a_{4}}}{H_{p-4}^{1}} \label{kp11}
    \end{eqnarray}

    \begin{eqnarray}
    && \Pi_{bosonic,anti-brane,p-5}\approx \frac{ i^{2(p-N+1)} \Sigma_{L=0}^{N} \Sigma_{H=0}^{N-L}\Sigma_{a_{1}..a_{L}=0}^{p}
    \Sigma_{j_{1}..j_{H}=p+1}^{M}\frac{\lambda^{2}}{1.2...N}(X^{j_{1}}..X^{j_{H}})^{2}\partial_{a_{6}}..\partial_{a_{L}}
    \bar{\tilde{R}}^{a_{1},a_{2},a_{3},a_{4}}}{H_{p-4}^{1}} \label{kp12}
    \end{eqnarray}

    \begin{eqnarray}
    && \Pi_{fermionic,anti-brane,p-5}\approx \nonumber \\ &&
    \frac{ -i^{2(p-N+1)+1} \Sigma_{L=0}^{N} \Sigma_{H=0}^{N-L}\Sigma_{a_{1}..a_{L}=0}^{p}
    \Sigma_{j_{1}..j_{H}=p+1}^{M}\frac{\lambda^{2}}{1.2...N}(X^{j_{1}}..X^{j_{H}})^{2}\partial_{a_{6}}..\partial_{a_{L}}
    \hat{\bar{R}}^{a_{1},a_{2},a_{3},a_{4}}}{H_{p-4}^{1}} \label{kp13}
    \end{eqnarray}

 We replace derivatives of order $p-5$ by these momentums and  obtain the Hamiltonian as follows:

     \begin{eqnarray}
      &&
      H_{p-5}^{1}\approx 4\pi\int d\sigma \sigma^{p-1}\Sigma_{L=0}^{N} \Sigma_{H=0}^{N-L}\Sigma_{a_{1}..a_{L}=0}^{p}
      \Sigma_{j_{1}..j_{H}=p+1}^{M}\frac{\lambda^{2}}{1.2...N}(X^{j_{1}}..X^{j_{H}})^{2}\partial_{a_{5}}..\partial_{a_{L}}
      \tilde{R}^{a_{1},a_{2},a_{3},a_{4}} \Pi_{bosonic,brane,p-5}i^{2(p-N)}\times\nonumber\\&&\sqrt{1+(\frac{i^{2(p-N)}
      k_{bosonic,brane,p-4}}{\sigma^{p-1}})^{2}} \nonumber\\&&
     -4\pi\int d\sigma \sigma^{p-1} \Sigma_{L=0}^{N} \Sigma_{H=0}^{N-L}\Sigma_{a_{1}..a_{L}=0}^{p}\Sigma_{j_{1}..j_{H}=p+1}^{M}
     \frac{\lambda^{2}}{1.2...N}(X^{j_{1}}..X^{j_{H}})^{2}\partial_{a_{5}}..\partial_{a_{L}} \bar{R}^{a_{1},a_{2},a_{3},a_{4}}
     \Pi_{fermionic,brane,p-5}i^{2(p-N)+1}\times\nonumber\\&&\sqrt{1+(\frac{i^{2(p-N)+1}k_{fermionic,brane,p-4}}{\sigma^{p-1}})^{2}}
     \nonumber\\&& +4\pi\int d\sigma \sigma^{p-1}\Sigma_{L=0}^{N} \Sigma_{H=0}^{N-L}\Sigma_{a_{1}..a_{L}=0}^{p}
     \Sigma_{j_{1}..j_{H}=p+1}^{M}\frac{\lambda^{2}}{1.2...N}(X^{j_{1}}..X^{j_{H}})^{2}\partial_{a_{5}}..\partial_{a_{L}}
     \hat{\tilde{R}}^{a_{1},a_{2},a_{3},a_{4}} \Pi_{bosonic,anti-brane,p-5}i^{2(p-N+1)}\times\nonumber\\&&
     \sqrt{1+(\frac{i^{2(p-N+1)}k_{bosonic,anti-brane,p-4}}{\sigma^{p-1}})^{2}} \nonumber\\&&+4\pi\int d\sigma
     \Sigma_{L=0}^{N} \Sigma_{H=0}^{N-L}\Sigma_{a_{1}..a_{L}=0}^{p}\Sigma_{j_{1}..j_{H}=p+1}^{M}
     \frac{\lambda^{2}}{1.2...N}(X^{j_{1}}..X^{j_{H}})^{2}\partial_{a_{5}}..\partial
 _{a_{L}} \hat{\bar{R}}^{a_{1},a_{2},a_{3},a_{4}}\Pi_{fermionic,anti-brane,p-5}i^{2(p-N+1)+1}\times\nonumber\\&&
 \sqrt{1+(\frac{i^{2(p-N+1)+1}k_{fermionic,anti-brane,p-4}}{\sigma^{p-1}})^{2}} \nonumber\\&&-4\pi\int d\sigma
 \sigma^{p-1}\Sigma_{L=0}^{N} \Sigma_{H=0}^{N-L}\Sigma_{a_{1}..a_{L}=0}^{p}\Sigma_{j_{1}..j_{H}=p+1}^{M}
 \frac{\lambda^{2}}{1.2...N}(X^{j_{1}}..X^{j_{H}})^{2}\partial_{a_{6}}..\partial_{a_{L}} \tilde{R}^{a_{1},a_{2},a_{3},a_{4}}
 \times\nonumber\\&&\partial_{a_{6}}(\sqrt{1+(\frac{i^{2(p-N)}k_{bosonic,brane,p-4}}{\sigma^{p-1}})^{2}}\sigma^{p-1}
 \Pi_{bosonic,brane,p-5}i^{2(p-N)}) \nonumber\\&&
  -4\pi\int d\sigma  \Sigma_{L=0}^{N} \Sigma_{H=0}^{N-L}\Sigma_{a_{1}..a_{L}=0}^{p}\Sigma_{j_{1}..j_{H}=p+1}^{M}
  \frac{\lambda^{2}}{1.2...N}(X^{j_{1}}..X^{j_{H}})^{2}\partial_{a_{6}}..\partial_{a_{L}} \bar{R}^{a_{1},a_{2},a_{3},a_{4}}
  \times\nonumber\\&&\partial_{a_{6}}(\sqrt{1+(\frac{i^{2(p-N)+1}k_{fermionic,brane,p-4}}{\sigma^{p-1}})^{2}}\sigma^{p-1}
  \Pi_{fermionic,brane,p-5}i^{2(p-N)+1} )\nonumber\\&& -4\pi\int d\sigma \Sigma_{L=0}^{N} \Sigma_{H=0}^{N-L}
  \Sigma_{a_{1}..a_{L}=0}^{p}\Sigma_{j_{1}..j_{H}=p+1}^{M}\frac{\lambda^{2}}{1.2...N}(X^{j_{1}}..X^{j_{H}})^{2}
  \partial_{a_{6}}..\partial_{a_{L}} \hat{\tilde{R}}^{a_{1},a_{2},a_{3},a_{4}}\times\nonumber\\&& \partial_{a_{6}}
  (\sqrt{1+(\frac{i^{2(p-N+1)}k_{bosonic,anti-brane,p-4}}{\sigma^{p-1}})^{2}}\sigma^{p-1}
  \Pi_{bosonic,anti-brane,p-5}i^{2(p-N+1)}) \nonumber\\&&+4\pi\int d\sigma  \Sigma_{L=0}^{N} \Sigma_{H=0}^{N-L}
  \Sigma_{a_{1}..a_{L}=0}^{p}\Sigma_{j_{1}..j_{H}=p+1}^{M}\frac{\lambda^{2}
 }{1.2...N}(X^{j_{1}}..X^{j_{H}})^{2}\partial_{a_{6}}..\partial_{a_{L}} \hat{\bar{R}}^{a_{1},a_{2},a_{3},a_{4}}\times
 \nonumber\\&&\partial_{a_{6}}(\sqrt{1+(\frac{i^{2(p-N+1)+1}k_{fermionic,anti-brane,p-4}}{\sigma^{p-1}})^{2}}\sigma^{p-1}
 \Pi_{fermionic,anti-brane,p-5}i^{2(p-N+1)+1})  \nonumber\\&&-L_{bosonic,brane,p-4}^{1}-L_{fermionic,brane,p-4}^{1}\nonumber\\&&
 -L_{bosonic,anti-brane,p-4}^{1}-L_{fermionic,anti-brane,p-4}^{1}
    \label{kp14}
   \end{eqnarray}
 Similar to previous stage, we use the constraints ($\partial_{a_{6}}(\sqrt{1+(\frac{i^{.....}k}{\sigma^{p-1}})^{2}}
 \sigma^{p-1}\Pi i^{..})=0 $) and obtain the following momentums:

   \begin{eqnarray}
   && \Pi_{bosonic,brane,p-5}=\frac{i^{2(p-N)}k_{bosonic,brane,p-5}}{\sigma^{p-1}\sqrt{1+(\frac{i^{2(p-N)}k_{bosonic,brane,p-4}}
   {\sigma^{p-1}})^{2}}}\nonumber\\&&\Pi_{fermionic,brane,p-4}=-\frac{i^{2(p-N)+1}k_{fermionic,brane,p-4}}{\sigma^{p-1}
   \sqrt{1+(\frac{i^{2(p-N)+1}k_{fermionic,brane,p-4}}{\sigma^{p-1}})^{2}}}\nonumber\\&& \Pi_{bosonic,anti-brane,p-5}=
   \frac{i^{2(p-N+1)}k_{bosonic,anti-brane,p-5}}{\sigma^{p-1}\sqrt{1+(\frac{i^{2(p-N+1)}k_{bosonic,anti-brane,p-4}}
   {\sigma^{p-1}})^{2}}}\nonumber\\&&\Pi_{fermionic,anti-brane,p-5}=-\frac{i^{2(p-N+1)+1}k_{fermionic,anti-brane,p-5}}
   {\sigma^{p-1}\sqrt{1+(\frac{i^{2(p-N+1)+1}k_{fermionic,anti-brane,p-4}}{\sigma^{p-1}})^{2}}}
   \label{kp15}
    \end{eqnarray}

  Substituting these momentums in Hamiltonian (\ref{kp14}), we derive the following Hamiltonian:

    \begin{eqnarray}
      &&
    H_{p-4}^{1}\approx 4\pi\int d\sigma \sigma^{p-1}\Big([ \frac{1}{2}i^{2(p-N)}\Sigma_{L=0}^{N} \Sigma_{H=0}^{N-L}
    \Sigma_{a_{1}..a_{L}=0}^{p}\Sigma_{j_{1}..j_{H}=p+1}^{M}(X^{j_{1}}..X^{j_{H-1}})^{2}\langle \partial_{a_{1}}..
    \partial_{a_{L}}X^{i},\partial_{a_{1}}..\partial_{a_{L}}X^{i}\rangle + \nonumber \\
       &&
   i^{2(p-N)} \Sigma_{L=0}^{N} \Sigma_{H=0}^{N-L}\Sigma_{a_{1}..a_{L}=0}^{p}\Sigma_{j_{1}..j_{H}=p+1}^{M}
   \frac{\lambda^{2}}{1.2...N}(X^{j_{1}}..X^{j_{H}})^{2}\langle \partial_{a_{7}}..\partial_{a_{L}}
   \tilde{R}^{a_{1},a_{2},a_{3},a_{4}},\partial_{a_{7}}..\partial_{a_{L}} \tilde{R}^{a_{1},a_{2},a_{3},a_{4}}
   \rangle+... ]^{1/2}\times\nonumber \\
    &&\sqrt{1+(\frac{i^{2(p-N)}k_{bosonic,brane,p-5}}{\sigma^{p-1}\sqrt{1+(\frac{i^{2(p-N)}k_{bosonic,brane,p-4}}
    {\sigma^{p-1}})^{2}}})^{2}}+\nonumber \\&&
    [-\frac{1}{2}i^{2(p-N)+1}\Sigma_{L=0}^{N} \Sigma_{H=0}^{N-L}\Sigma_{a_{1}..a_{L}=0}^{p}\Sigma_{j_{1}..j_{H}=p+1}^{M}
    (X^{j_{1}}..X^{j_{H-1}})^{2}\gamma^{a_{L-1}}\langle \partial_{a_{1}}..\partial_{a_{L-1}}\psi^{i},\partial_{a_{1}}..
    \partial_{a_{L}}\psi^{i}\rangle -\nonumber \\
    &&  i^{2(p-N)+1}\Sigma_{L=0}^{N} \Sigma_{H=0}^{N-L}\Sigma_{a_{1}..a_{L}=0}^{p}\Sigma_{j_{1}..j_{H}=p+1}^{M}
    \frac{\lambda^{2}}{1.2...N}\gamma^{a_{L-1}}(X^{j_{1}}..X^{j_{H}})^{2}\langle \partial_{a_{7}}..\partial_{a_{L-1}}
    \bar{R}^{a_{1},a_{2},a_{3},a_{4}},\partial_{a_{7}}..\partial_{a_{L}} \bar{R}^{a_{1},a_{2},a_{3},a_{4}}\rangle +...]^{1/2}
    \times  \nonumber \\&& \sqrt{1+(-\frac{i^{2(p-N)+1}k_{fermionic,brane,p-5}}{\sigma^{p-1}
    \sqrt{1+(-\frac{i^{2(p-N)+1}k_{fermionic,brane,p-4}}{\sigma^{p-1}})^{2}}})^{2}}+ \nonumber \\&& [ \frac{1}{2}i^{2(p-N+1)}
    \Sigma_{L=0}^{N} \Sigma_{H=0}^{N-L}\Sigma_{a_{1}..a_{L}=0}^{p}\Sigma_{j_{1}..j_{H}=p+1}^{M}(X^{j_{1}}..X^{j_{H-1}})^{2}
    \langle \partial_{a_{1}}..\partial_{a_{L}}X^{i},\partial_{a_{1}}..\partial_{a_{L}}X^{i}\rangle + \nonumber \\
       &&
   i^{2(p-N)} \Sigma_{L=0}^{N} \Sigma_{H=0}^{N-L}\Sigma_{a_{1}..a_{L}=0}^{p}\Sigma_{j_{1}..j_{H}=p+1}^{M}
   \frac{\lambda^{2}}{1.2...N}(X^{j_{1}}..X^{j_{H}})^{2}\langle \partial_{a_{6}}..\partial_{a_{L}}
   \hat{\tilde{R}}^{a_{1},a_{2},a_{3},a_{4}},\partial_{a_{6}}..\partial_{a_{L}} \hat{\tilde{R}}^{a_{1},a_{2},a_{3},a_{4}}
   \rangle+... ]^{1/2}\times\nonumber \\&&\sqrt{1+(\frac{i^{2(p-N+1)}k_{bosonic,anti-brane,p-5}}{\sigma^{p-1}
   \sqrt{1+(\frac{i^{2(p-N+1)}k_{bosonic,anti-brane,p-4}}{\sigma^{p-1}})^{2}}})^{2}} +\nonumber \\&&
   [  - \frac{1}{2}i^{2(p-N+1)+1}\Sigma_{L=0}^{N} \Sigma_{H=0}^{N-L}\Sigma_{a_{1}..a_{L}=0}^{p}\Sigma_{j_{1}..j_{H}=p+1}^{M}(X^{j_{1}}..X^{j_{H-1}})^{2}
   \gamma^{a_{L-1}}\langle \partial_{a_{1}}..\partial_{a_{L-1}}\psi^{i},\partial_{a_{1}}..\partial_{a_{L}}\psi^{i}\rangle -\nonumber \\
   &&  i^{2(p-N)+1}\Sigma_{L=0}^{N} \Sigma_{H=0}^{N-L}\Sigma_{a_{1}..a_{L}=0}^{p}\Sigma_{j_{1}..j_{H}=p+1}^{M}
   \frac{\lambda^{2}}{1.2...N}\gamma^{a_{L-1}}(X^{j_{1}}..X^{j_{H}})^{2}\langle \partial_{a_{6}}..
   \partial_{a_{L-1}}\hat{\bar{R}}^{a_{1},a_{2},a_{3},a_{4}},\partial_{a_{6}}..\partial_{a_{L}}
   \hat{\bar{R}}^{a_{1},a_{2},a_{3},a_{4}}\rangle +..]^{1/2}\times\nonumber \\
   &&\sqrt{1+(-\frac{i^{2(p-N+1)+1}k_{fermionic,anti-brane,p-5}}{\sigma^{p-1}\sqrt{1+(-\frac{i^{2(p-N+1)+1}k_{fermionic,anti-brane,p-4}}
   {\sigma^{p-1}})^{2}}})^{2}}\Big)
    \label{kp16}
    \end{eqnarray}

After doing some mathematical calculations, we can remove all derivatives respect to curvatures and obtain the Hamiltonian as follows:

    \begin{eqnarray}
     &&
  H_{tot}^{1}\approx 4\pi\int d\sigma \sigma^{p-1} \Big([ \frac{1}{2}i^{2(p-N)}\Sigma_{L=0}^{N} \Sigma_{H=0}^{N-L}
  \Sigma_{a_{1}..a_{L}=0}^{p}\Sigma_{j_{1}..j_{H}=p+1}^{M}(X^{j_{1}}..X^{j_{H-1}})^{2}\langle \partial_{a_{1}}..
   \partial_{a_{L}}X^{i},\partial_{a_{1}}..\partial_{a_{L}}X^{i}\rangle + ... ]^{1/2}\times\nonumber \\
    &&F_{bosonic,brane,tot}+\nonumber \\&&
  [-\frac{1}{2}i^{2(p-N)+1}\Sigma_{L=0}^{N} \Sigma_{H=0}^{N-L}\Sigma_{a_{1}..a_{L}=0}^{p}\Sigma_{j_{1}..j_{H}=p+1}^{M}
  (X^{j_{1}}..X^{j_{H-1}})^{2}\gamma^{a_{L-1}}\langle \partial_{a_{1}}..\partial_{a_{L-1}}\psi^{i},\partial_{a_{1}}..
  \partial_{a_{L}}\psi^{i}\rangle +...]^{1/2} \times  \nonumber \\&&F_{fermionic,brane,tot}+ \nonumber \\&&
  [ \frac{1}{2}i^{2(p-N+1)}\Sigma_{L=0}^{N} \Sigma_{H=0}^{N-L}\Sigma_{a_{1}..a_{L}=0}^{p}\Sigma_{j_{1}..j_{H}=p+1}^{M}
  (X^{j_{1}}..X^{j_{H-1}})^{2}\langle \partial_{a_{1}}..\partial_{a_{L}}X^{i},\partial_{a_{1}}..\partial_{a_{L}}X^{i}
  \rangle +... ]^{1/2}\times\nonumber \\&&F_{bosonic,anti-brane,tot}  +\nonumber \\&&
  [  - \frac{1}{2}i^{2(p-N+1)+1}\Sigma_{L=0}^{N} \Sigma_{H=0}^{N-L}\Sigma_{a_{1}..a_{L}=0}^{p}\Sigma_{j_{1}..j_{H}=p+1}^{M}
  (X^{j_{1}}..X^{j_{H-1}})^{2}\gamma^{a_{L-1}}\langle \partial_{a_{1}
 }..\partial_{a_{L-1}}\psi^{i},\partial_{a_{1}}..\partial_{a_{L}}\psi^{i}\rangle +..]^{1/2}\times\nonumber \\
 &&F_{fermionic,anti-brane,tot}\Big)
    \label{kp17}
  \end{eqnarray}
 where functions of $F$  are defined as follows:

  \begin{eqnarray}
  && F_{bosonic,brane,tot}= \sqrt{1+(\frac{i^{2(p-N)}k_{bosonic,brane,1}}{\sigma^{p-1}\sqrt{1+(\frac{i^{2(p-N)}k_{bosonic,brane,2}}
  {\sigma^{p-1}\sqrt{1+(\frac{i^{2(p-N)}k_{bosonic,brane,3}}{\sigma^{p-1}...\sqrt{1+(\frac{i^{2(p-N)}k_{bosonic,brane,p-4}}
  {\sigma^{p-1}})^{2}}})^{2}}})^{2}}})^{2}}\nonumber \\ &&F_{fermionic,brane,tot}= \sqrt{1+(\frac{i^{2(p-N)+1}k_{fermionic,brane,1}}
  {\sigma^{p-1}\sqrt{1+(\frac{i^{2(p-N)+1}k_{fermionic,brane,2}}{\sigma^{p-1}\sqrt{1+(\frac{i^{2(p-N)+1}k_{fermionic,brane,3}}
  {\sigma^{p-1}...\sqrt{1+(\frac{i^{2(p-N)+1}k_{fermionic,brane,p-4}}{\sigma^{p-1}})^{2}}})^{2}}})^{2}}})^{2}}\nonumber \\
   &&F_{bosonic,anti-brane,tot}= \sqrt{1+(\frac{i^{2(p-N+1)}k_{bosonic,anti-brane,1}}{\sigma^{p-1}
   \sqrt{1+(\frac{i^{2(p-N+1)}k_{bosonic,anti-brane,2}}{\sigma^{p-1}\sqrt{1+(\frac{i^{2(p-N+1)}k_{bosonic,anti-brane,3}}
   {\sigma^{p-1}...\sqrt{1+(\frac{i^{2(p-N+1)}k_{bosonic,anti-brane,p-4}}{\sigma^{p-1}})^{2}}})^{2}}})^{2}}})^{2}}\nonumber \\
   &&F_{fermionic,anti-brane,tot}= \sqrt{1+(\frac{i^{2(p-N+1)+1}k_{fermionic,anti-brane,1}}{\sigma^{p-1}
   \sqrt{1+(\frac{i^{2(p-N+1)+1}k_{fermionic,anti-brane,2}}{\sigma^{p-1}\sqrt{1+(\frac{i^{2(p-N+1)+1}k_{fermionic,anti-brane,3}}
   {\sigma^{p-1}...\sqrt{1+(\frac{i^{2(p-N+1)+1}k_{fermionic,anti-brane,p-4}}{\sigma^{p-1}})^{2}}})^{2}}})^{2}}})^{2}}
   \label{kp18}
   \end{eqnarray}

   These results are the very same as Hamiltonians of BIon in \cite{k7,k8,k23}. It is clear from the above equations that
   curvatures of bosonic gravitons and fermionic gravitinoes produce two types of wormholes in which their signatures and
   couplings are different. These wormholes can act against each other and also cancel the effect of each other. In addition to that,
   the sign of Hamiltonians of wormholes which are created by bosonic gravitons and fermionic gravitinoes on anti-branes
   is opposite. This means that the potential energy of one brane has negative sign and it attracts particles and the
   potential energy of another brane has positive sign and it repels particles and thus particles move from one brane
   to another and a wormhole is formed between two branes.  Now, we simplify calculations by choosing
   $x^{0}=it,x^{1,2,3}=\sigma$, $X^{0}=t$,$X^{1}=z_{+}+iz_{-}$, $X^{i}=0,i\neq 0,1$, $\psi^{0}=t$,  $\psi^{1}=y_{+}+iy_{-}$,
   $\psi^{i}=0,i\neq 0,1$, $i^{2(p-N)}=1$ and $\gamma^{a_{L-1}}=i^{2(N-L-1)-1}$ where indices $\pm$ denote the fields
   on brane and anti-branes respectively. Also, $\sigma$ denotes the separation between quarks on one brane, z and y refer
   to lengths of bosonic and fermionic wormholes between branes and their $n^{th}$ derivatives are shown by $z^{n(')},y^{n(')}$.
   Putting these assumption in Eq. (\ref{kp17}), we rewrite Hamiltonian as:

      \begin{eqnarray}
       &&
      H_{tot}^{1}\approx 4\pi\int d\sigma \sigma^{p-1} \Big([1+ \Sigma_{n=1}^{N-1}(z_{+}^{(N-n-1)}z_{+}^{n(')})^{2} ]^{1/2}
      F_{bosonic,brane,tot}+\nonumber \\&&
       [-1+\Sigma_{n=1}^{N-1}(-iz_{+})^{2(N-n-1)}y_{+}^{n}y_{+}^{n(')}]^{1/2} F_{fermionic,brane,tot}+ \nonumber \\&&
       [-1+ i^{2N-2}\Sigma_{n=1}^{N-1}(z_{-}^{(N-n-1)}z_{-}^{n(')})^{2} ]^{1/2}F_{bosonic,anti-brane,tot} +\nonumber \\&&
       [1+i^{2N-2}\Sigma_{n=1}^{N-1}(-iz_{-})^{2(N-n-1)}y_{-}^{n}y_{-}^{n(')}]^{1/2}F_{fermionic,anti-brane,tot}\Big)
      \label{kp19}
     \end{eqnarray}

   Now, we can obtain the equation of motion for $z$ and $y$:

    \begin{eqnarray}
    &&
   \Big(\Sigma_{n=1}^{N-1}(-1)^{n}(z_{+}^{(N-n-1)})^{2}z_{+}^{(n)(')}z_{+}^{(n-1)(')}\frac{\sigma^{p-1}F_{bosonic,brane,tot}}
   {[1+ \Sigma_{n=1}^{N-1}(z_{+}^{(N-n-1)}z_{+}^{n(')})^{2} ]^{1/2}}\Big)'=\nonumber\\&&
   \Big(\Sigma_{n=1}^{N-1}(z_{+}^{n(')})^{2}z_{+}^{2(N-n-1)-1}\frac{\sigma^{p-1}F_{bosonic,brane,tot}}
   {[1+ \Sigma_{n=1}^{N-1}(z_{+}^{(N-n-1)}z_{+}^{n(')})^{2} ]^{1/2}}\Big)
    \label{kp20}
   \end{eqnarray}

   \begin{eqnarray}
    &&
  \Big(\Sigma_{n=1}^{N-1}(-1)^{n}(-iz_{+})^{2(N-n-1)}y_{+}^{n}y_{+}^{(n-1)(')}\frac{\sigma^{p-1}F_{fermionic,brane,tot}}
  { [-1+\Sigma_{n=1}^{N-1}(-iz_{+})^{2(N-n-1)}y_{+}^{n}y_{+}^{n(')}]^{1/2}}\Big)'=\nonumber\\&&\Big(\Sigma_{n=1}^{N-1}
  (-iz_{+})^{2(N-n-1)}y_{+}^{n-1}y_{+}^{n(')}\frac{\sigma^{p-1}F_{fermionic,brane,tot}}
  { [-1+\Sigma_{n=1}^{N-1}(-iz_{+})^{2(N-n-1)}y_{+}^{n}y_{+}^{n(')}]^{1/2}}\Big)
  \label{kp21}
   \end{eqnarray}

   \begin{eqnarray}
   &&
  \Big(\Sigma_{n=1}^{N-1}i^{2N-2}(-1)^{n}(z_{-}^{(N-n-1)})^{2}z_{-}^{(n)(')}z_{-}^{(n-1)(')}\frac{\sigma^{p-1}
  F_{bosonic,anti-brane,tot}}{[-1+ i^{2N-2}\Sigma_{n=1}^{N-1}(z_{-}^{(N-n-1)}z_{-}^{n(')})^{2} ]^{1/2}}\Big)'=\nonumber\\&&
  \Big(\Sigma_{n=1}^{N-1}i^{2N-2}(z_{-}^{n(')})^{2}z_{+}^{2(N-n-1)-1}\frac{\sigma^{p-1}F_{bosonic,anti-brane,tot}}{[-1+ i^{2N-2}
  \Sigma_{n=1}^{N-1}(z_{-}^{(N-n-1)}z_{-}^{n(')})^{2} ]^{1/2} }\Big)
  \label{kp22}
  \end{eqnarray}

   \begin{eqnarray}
   &&
  \Big(\Sigma_{n=1}^{N-1}i^{2N-2}(-1)^{n}(-iz_{-})^{2(N-n-1)}y_{-}^{n}y_{-}^{(n-1)(')}\frac{\sigma^{p-1}F_{fermionic,anti-brane,tot}}
  { [1+i^{2N-2}\Sigma_{n=1}^{N-1}(-iz_{-})^{2(N-n-1)}y_{-}^{n}y_{-}^{n(')}]^{1/2}}\Big)'=\nonumber\\&&
  \Big(\Sigma_{n=1}^{N-1}i^{2N-2}(-iz_{-})^{2(N-n-1)}y_{-}^{n-1}y_{-}^{n(')}\frac{\sigma^{p-1}F_{fermionic,anti-brane,tot}}
  { [1+i^{2N-2}\Sigma_{n=1}^{N-1}(-iz_{-})^{2(N-n-1)}y_{-}^{n}y_{-}^{n(')}]^{1/2}}\Big)
  \label{kp23}
   \end{eqnarray}

  Solving these equations, we obtain:

  \begin{eqnarray}
  && z_{+}=z_{+,0}\Sigma_{n=0}^{N-1}e^{-\int d^{n}\sigma F_{bosonic,brane,tot}(\sigma)\frac{1}{F_{bosonic,brane,tot}^{-n}
  (\sigma_{0,bosonic,brane})-F_{bosonic,brane,tot}^{-n}(\sigma)}}  \Big(1+\nonumber\\&&
  \int d^{n}\sigma F_{bosonic,brane,tot}^{-1}(\sigma)(F_{bosonic,brane,tot}^{-n}(\sigma_{0,bosonic,brane})-
  F_{bosonic,brane,tot}^{-n}(\sigma))\sin(n\sigma)  \Big)
  \label{kp24}
  \end{eqnarray}

   \begin{eqnarray}
   && y_{+}=y_{+,0}\Sigma_{n=0}^{N-1}e^{-\int d^{n}\sigma F_{fermionic,brane,tot}^{-1}(\sigma)\frac{1}{F_{fermionic,brane,tot}^{n}
   (\sigma)-F_{fermionic,brane,tot}^{n}(\sigma_{0,fermionic,brane})}}  \Big(1+\nonumber\\&&
   \int d^{n}\sigma F_{fermionic,brane,tot}(\sigma)(F_{fermionic,brane,tot}^{n}(\sigma)-F_{fermionic,brane,tot}^{n}
   (\sigma_{0,fermionic,brane}))\cos(n\sigma)  \Big)\label{kp25}
   \end{eqnarray}

     \begin{eqnarray}
  && z_{-}=z_{-,0}\Sigma_{n=0}^{N-1}e^{-\int d^{n}\sigma F_{bosonic,anti-brane,tot}(\sigma)\frac{1}{F_{bosonic,anti-brane,tot}^{-n}
  (\sigma)-F_{bosonic,anti-brane,tot}^{-n}(\sigma_{0,bosonic,anti-brane})}}  \Big(1+\nonumber\\&&
  \int d^{n}\sigma F_{bosonic,anti-brane,tot}^{-1}(\sigma)(F_{bosonic,anti-brane,tot}^{-n}(\sigma)- \nonumber\\&&
  F_{bosonic,anti-brane,tot}^{-n}
  (\sigma_{0,bosonic,anti-brane}))\sin(n\sigma)  \Big)\label{kp26}
    \end{eqnarray}

    \begin{eqnarray}
    && y_{-}=y_{-,0}\Sigma_{n=0}^{N-1}e^{-\int d^{n}\sigma F_{fermionic,anti-brane,tot}^{-1}(\sigma)
    \frac{1}{F_{fermionic,anti-brane,tot}^{n}(\sigma_{0,fermionic,anti-brane})-F_{bosonic,anti-brane,tot}^{n}(\sigma)}}
    \Big(1+\nonumber\\&&
    \int d^{n}\sigma F_{fermionic,anti-brane,tot}(\sigma)(F_{fermionic,anti-brane,tot}^{n}(\sigma_{0,fermionic,anti-brane})-\nonumber\\&&
    F_{fermionic,anti-brane,tot}^{n}(\sigma))\cos(n\sigma)\Big),
    \label{kp27}
     \end{eqnarray}
 where  $\sigma_{0}$ is throat of wormhole. To construct quarkonium in BIon, we assume that ($\sigma$) is the separation
 distance between quarks and anti-quarks. Thus, these solutions show that at $\sigma=0$, the length of gravitonic wormholes
 is zero, by increasing the separation distance between two quarks ($\sigma$), this length grows, turns over a maximum and
 reduces to zero at throat $\sigma_{0}$ and then one new bosonic wormhole is born, it's length increases  with increasing
 $\sigma$ and tends to infinity at $\sigma=\infty$. On the other hand, the length of fermionic wormholes is $\infty$ at $\sigma=0$
 and reduces to zero at throat $\sigma_{0}$, then one new fermionic wormhole is formed, it's length grows with increasing $\sigma$,
 turns over a maximum and reduces to zero at $\infty$. Thus, fermionic and bosonic wormholes act against to each other and this
 prevents the closing in and getting away of quarks from each other. \\

 % Using Eqs. (\ref{kp24},\ref{kp25},\ref{kp26},\ref{kp27}) in Eq. (\ref{kp19}), we obtain the potential for this system:

 Using Eqs. (\ref{kp24}-\ref{kp27}) in Eq. (\ref{kp19}), we obtain the potential for this system:

\begin{eqnarray}
   && H_{tot}\approx V_{tot}=V_{bosonic,brane,tot}+V_{fermionic,brane,tot}+V_{bosonic,anti-brane,tot}+V_{fermionic,anti-brane,tot} \label{kp28}
\end{eqnarray}
\begin{eqnarray}
	&&
    V_{bosonic,brane,tot}\approx 4\pi\int d\sigma \Big(\Big( \Sigma_{n'=1}^{N-1}[z_{+,0}\Sigma_{n=0}^{N-1}e^{-\int d^{n}\sigma F_{bosonic,brane,tot}
    (\sigma)\frac{1}{F_{bosonic,brane,tot}^{-n}(\sigma_{0,bosonic,brane})-F_{bosonic,brane,tot}^{-n}(\sigma)}} \nonumber\\&& \Big(1+
   \int d^{n}\sigma F_{bosonic,brane,tot}^{-1}(\sigma)(F_{bosonic,brane,tot}^{-n}(\sigma_{0,bosonic,brane})-F_{bosonic,brane,tot}^{-n}
   (\sigma))\sin(n\sigma)\Big)]^{2N-2n'-2}\times
   \nonumber\\&&[F_{bosonic,brane,tot}^{-1}(\sigma)(F_{bosonic,brane,tot}^{-n}(\sigma_{0,bosonic,brane})-F_{bosonic,brane,tot}^{-n}(\sigma))\sin(n\sigma)\times
   \nonumber\\&& e^{-\int d^{n}\sigma F_{bosonic,brane,tot}(\sigma)\frac{1}{F_{bosonic,brane,tot}^{-n}(\sigma_{0,bosonic,brane})-F_{bosonic,brane,tot}^{-n}(\sigma)}}+
   \nonumber\\&&e^{-\int d^{n}\sigma F_{bosonic,brane,tot}(\sigma)\frac{1}{F_{bosonic,brane,tot}^{-n}(\sigma_{0,bosonic,brane})-F_{bosonic,brane,tot}^{-n}
   (\sigma)}}\times\nonumber\\&&(F_{bosonic,brane,tot}(\sigma)\frac{1}{F_{bosonic,brane,tot}^{-n}(\sigma_{0,bosonic,brane})-F_{bosonic,brane,tot}^{-n}
   (\sigma)})]^{2}F_{bosonic,brane,tot}(\sigma) \Big)\Big)
\label{kp1t28}
\end{eqnarray}

\begin{eqnarray}
    && V_{fermionic,brane,tot}\approx 4\pi\int d\sigma \Big(\Big( \Sigma_{n'=1}^{N-1}[z_{+,0}\Sigma_{n=0}^{N-1}e^{-\int d^{n}\sigma F_{bosonic,brane,tot}
    (\sigma)\frac{1}{F_{bosonic,brane,tot}^{-n}(\sigma_{0,bosonic,brane})-F_{bosonic,brane,tot}^{-n}(\sigma)}} \nonumber\\&& \Big(1+
    \int d^{n}\sigma F_{bosonic,brane,tot}^{-1}(\sigma)(F_{bosonic,brane,tot}^{-n}(\sigma_{0,bosonic,brane})-F_{bosonic,brane,tot}^{-n}(\sigma))
    \sin(n\sigma)\Big) ]^{2N-2n'-2}\times\nonumber\\&&[y_{+,0}\Sigma_{n=0}^{N-1}e^{-\int d^{n}\sigma F_{fermionic,brane,tot}^{-1}(\sigma)
    \frac{1}{F_{fermionic,brane,tot}^{n}(\sigma)-F_{fermionic,brane,tot}^{n}(\sigma_{0,fermionic,brane})}}  \Big(1+\nonumber\\&&
    \int d^{n}\sigma F_{fermionic,brane,tot}(\sigma)(F_{fermionic,brane,tot}^{n}(\sigma)-F_{fermionic,brane,tot}^{n}(\sigma_{0,fermionic,brane}))
    \cos(n\sigma)  \Big)]^{n'}\times\nonumber\\&&[y_{+,0}\Sigma_{n=0}^{N-1}(e^{-\int d^{n}\sigma F_{fermionic,brane,tot}^{-1}(\sigma)
    \frac{1}{F_{fermionic,brane,tot}^{n}(\sigma)-F_{fermionic,brane,tot}^{n}(\sigma_{0,fermionic,brane})}}\times  \nonumber\\&&
    F_{fermionic,brane,tot}(\sigma)(F_{fermionic,brane,tot}^{n}(\sigma)-F_{fermionic,brane,tot}^{n}(\sigma_{0,fermionic,brane}))\cos(n\sigma))+
    \nonumber\\&& \Big(e^{-\int d^{n}\sigma F_{fermionic,brane,tot}^{-1}(\sigma)\frac{1}{F_{fermionic,brane,tot}^{n}(\sigma)-F_{fermionic,brane,tot}^{n}
    (\sigma_{0,fermionic,brane})}}\times  \nonumber\\&&F_{fermionic,brane,tot}^{-1}(\sigma)\frac{1}{F_{fermionic,brane,tot}^{n}(\sigma)-F_{fermionic,brane,tot}^{n}
    (\sigma_{0,fermionic,brane})}) \Big)]F_{fermionic,brane,tot}\Big)\Big)\label{kp2t28}
\end{eqnarray}

\begin{eqnarray}
   && V_{bosonic,anti-brane,tot}\nonumber\\&& \approx 4\pi\int d\sigma \Big(\Big( \Sigma_{n'=1}^{N-1}[z_{-,0}\Sigma_{n=0}^{N-1}e^{-\int d^{n}
    \sigma F_{bosonic,anti-brane,tot}(\sigma)\frac{1}{F_{bosonic,anti-brane,tot}^{-n}(\sigma)-F_{bosonic,brane,tot}^{-n}(\sigma_{0,anti-bosonic,brane})}}
    \nonumber\\&& \Big(1+ \int d^{n}\sigma F_{bosonic,anti-brane,tot}^{-1}(\sigma)(F_{bosonic,anti-brane,tot}^{-n}(\sigma)-F_{bosonic,brane,tot}^{-n}
    (\sigma_{0,anti-bosonic,brane}))\sin(n\sigma)  \Big)]^{2N-2n'-2}\times
    \nonumber\\&&[F_{bosonic,anti-brane,tot}^{-1}(\sigma)(F_{bosonic,anti-brane,tot}^{-n}(\sigma)-F_{bosonic,anti-brane,tot}^{-n}(\sigma_{0,anti-bosonic,brane}))
    \sin(n\sigma)\times\nonumber\\&& e^{-\int d^{n}\sigma F_{bosonic,anti-brane,tot}(\sigma)\frac{1}{F_{bosonic,anti-brane,tot}^{-n}
    (\sigma)-F_{bosonic,anti-brane,tot}^{-n}(\sigma_{0,anti-bosonic,brane})}}+\nonumber\\&&e^{-\int d^{n}\sigma F_{bosonic,anti-brane,tot}(\sigma)
    \frac{1}{F_{bosonic,anti-brane,tot}^{-n}(\sigma)-F_{bosonic,anti-brane,tot}^{-n}(\sigma_{0,anti-bosonic,brane})}}\times\nonumber\\&&
    (F_{bosonic,anti-brane,tot}(\sigma)\frac{1}{F_{bosonic,anti-brane,tot}^{-n}(\sigma)-F_{bosonic,anti-brane,tot}^{-n}(\sigma_{0,anti-bosonic,brane})})]^{2}\times
    \nonumber\\&&F_{bosonic,anti-brane,tot}(\sigma) \Big)\Big)\label{kp3t28}
    \end{eqnarray}

    \begin{eqnarray}
&&
    V_{fermionic, anti-brane,tot}\approx \nonumber\\&& 4\pi\int d\sigma \Big(\Big( \Sigma_{n'=1}^{N-1}[z_{-,0}\Sigma_{n=0}^{N-1}e^{-\int d^{n}
    \sigma F_{bosonic,anti-brane,tot}(\sigma)\frac{1}{F_{bosonic,anti-brane,tot}^{-n}(\sigma)-F_{bosonic,anti-brane,tot}^{-n}(\sigma_{0,bosonic,anti-brane})}}
    \nonumber\\&& \Big(1+ \int d^{n}\sigma F_{bosonic,anti-brane,tot}^{-1}(\sigma)(F_{bosonic,anti-brane,tot}^{-n}(\sigma)-F_{bosonic,anti-brane,tot}^{-n}
    (\sigma_{0,bosonic,anti-brane}))\times\nonumber\\&&\sin(n\sigma)\Big) ]^{2N-2n'-2}[y_{-,0}\Sigma_{n=0}^{N-1}\times\nonumber\\&&e^{-\int d^{n}
    \sigma F_{fermionic,anti-brane,tot}^{-1}(\sigma)\frac{1}{F_{fermionic,anti-brane,tot}^{n}(\sigma_{0,fermionic,anti-brane})-F_{fermionic,anti-brane,tot}^{n}
    (\sigma)}}\nonumber\\&&  \Big(1+ \int d^{n}\sigma F_{fermionic,anti-brane,tot}(\sigma)(F_{fermionic,anti-brane,tot}^{n}(\sigma_{0,fermionic,anti-brane})-
    \nonumber\\&& F_{fermionic,anti-anti-brane,tot}^{n}(\sigma))\cos(n\sigma)  \Big)]^{n'}\times\nonumber\\&&[y_{-,0}\Sigma_{n=0}^{N-1}
    (e^{-\int d^{n}\sigma F_{fermionic,anti-brane,tot}^{-1}(\sigma)\frac{1}{F_{fermionic,anti-brane,tot}^{n}(\sigma_{0,fermionic,anti-brane})-
    F_{fermionic,anti-brane,tot}^{n}(\sigma)}}\times  \nonumber\\&&
    F_{fermionic,anti-brane,tot}(\sigma)(F_{fermionic,anti-brane,tot}^{n}(\sigma_{0,fermionic,anti-brane})-F_{fermionic,anti-brane,tot}^{n}(\sigma))
    \cos(n\sigma))+ \nonumber\\&& \Big(e^{-\int d^{n}\sigma F_{fermionic,anti-brane,tot}^{-1}(\sigma)\frac{1}{F_{fermionic,anti-brane,tot}^{n}
    (\sigma_{0,fermionic,anti-brane})-F_{fermionic,anti-brane,tot}^{n}(\sigma)}}\times  \nonumber\\&&F_{fermionic,anti-brane,tot}^{-1}(\sigma)
    \frac{1}{F_{fermionic,anti-brane,tot}^{n}(\sigma_{0,fermionic,anti-brane})-F_{fermionic,anti-brane,tot}^{n}(\sigma)}) \Big)]\times\nonumber\\&&
    F_{fermionic,anti-brane,tot}\Big)\Big)
     \label{kp4t28}
   \end{eqnarray}

 For simplicity, we assume that $\sigma_{0,bosonic,brane}=\sigma_{0,bosonic,anti-brane}$ and  $\sigma_{0 fermionic,brane}=
 \sigma_{0,fermionic,anti-brane}$ and obtain the  potentials and their relative forces between quarks and anti-quarks approximately:

   \begin{eqnarray}
   &&   V_{tot}=V_{bosonic,brane+anti-brane,tot}+V_{fermionic,brane+anti-brane,tot}\label{fff1}
   \end{eqnarray}
   \begin{eqnarray}
   &&
   V_{bosonic,brane+anti-brane,tot}\approx -\Sigma_{m=1}^{P-1}\Sigma_{n=1}^{N-1} [k_{bosonic,brane,}-k_{bosonic,anti-brane}]^{m}
   \sigma^{m}\Big(\sigma_{0,bosonic,brane}^{nm}-\sigma^{nm}\Big)\nonumber\\&&\nonumber\\&&
   \text{ For $\sigma \ll \sigma_{0,bosonic,brane}$} \quad F_{bosonic}\approx\Sigma_{m=1}^{P-1}\Sigma_{n=1}^{N-1}
   [k_{bosonic,brane,}-k_{bosonic,anti-brane}]^{m}m\sigma^{m-1}\sigma_{0,bosonic,brane}^{nm}\nonumber\\&&
   \text{ For $\sigma \gg \sigma_{0,bosonic,brane}$} \quad F_{bosonic}\approx-\Sigma_{m=1}^{P-1}
   \Sigma_{n=1}^{N-1} [k_{bosonic,brane,}-k_{bosonic,anti-brane}]^{m}n(m+1)\sigma^{m}\sigma^{nm-1}\label{fff2}
   \end{eqnarray}
   \begin{eqnarray}
   &&
   V_{fermionic,brane+anti-brane,tot}\approx \nonumber\\&&
   \Sigma_{m=1}^{P-1}\Sigma_{n=1}^{N-1} [k_{fermionic,brane,}-k_{fermionic,anti-brane}]^{m}
   \frac{1}{\sigma^{m}}\Big(\sigma^{-nm}-\sigma^{-nm}_{0,fermionic,brane}\Big) \nonumber\\&&\nonumber\\&&
   \text{ For $\sigma \ll \sigma_{0,fermionic,brane}$} \quad F_{fermionic}\approx \Sigma_{m=1}^{P-1}\Sigma_{n=1}^{N-1}
   [k_{fermionic,brane,}-k_{fermionic,anti-brane}]^{m}\frac{m+nm}{\sigma^{m+nm+1}}\nonumber\\&&
   \text{ For $\sigma \gg \sigma_{0,fermionic,brane}$} \quad F_{fermionic}\approx-\Sigma_{m=1}^{P-1}
   \Sigma_{n=1}^{N-1} [k_{fermionic,brane,}-k_{fermionic,anti-brane}]^{m} \times \nonumber\\&&
   \frac{m+1}{\sigma^{m+1}\sigma^{nm}_{0,fermionic,brane}}
   \label{fff3}
    \end{eqnarray}

 These results show that when quarks and anti-quarks are a very close ( $\sigma=0$), the potential of gravitonic wormholes is
 zero. By increasing the separation distance between these particles, bosonic wormhole produces a repulsive potential and anti-gravity
 force which first grows, turns over a maximum and then shrinks to zero at $\sigma_{0,bosonic,brane}$. After this distance,
 bosonic potential becomes attractive, gravity emerges and prevents the getting away of quarks from anti-quarks. On the other
 hand, the gravitino produces a wormhole which leads to creation of repulsive potential and  anti-gravity for small separation
 distance. This potential is $\infty $ at  ( $\sigma=0$) and causes the quarks and anti-quarks to get away from each other fastly.
 By increasing the separation distance between these particles, repulsive gravity decreases and shrinks to zero at
 $\sigma_{0,fermionic,brane}$. Then, the sign of potential reverses and anti-gravity changes to gravity. This gravity
 grows, turns over a maximum and shrinks to zero at $\infty $.
%%%%%%%%%%%%%%%%%%%%%%%%%%%%%%%%%%%%%%%%%%%%%%%%%%%%%%%%%%%%%%%%%%%%%%%%%%%%%%%%%%%%%%%%%%%%%%%%%%%%%
%%%%%%%%%%%%%%%%%%%%%%%%%%%%%%%%%%%% SECTION III %%%%%%%%%%%%%%%%%%%%%%%%%%%%%%%%%%%%%%%%%%%%%%%%
\section{Thermal quarkonium in a thermal BIon }\label{o2}
Until now, we have shown that for small separation distance between quarks and anti-quarks
($\sigma <\sigma_{0,bosonic/fermionic,brane}$), the  gravitational potentials which are produced by bosonic and fermionic
wormholes, are repulsive and thus, one repulsive force causes the getting away of particles from each other. However, for
large distance between quarks and anti-quarks ($\sigma >\sigma_{0,bosonic/fermionic,brane}$), the  gravitational potential
is attractive and thus, attractive force leads to closing particles toward each other.  In this section, we show that by
increasing temperature, the boundary between repulsive and attractive potential ($\sigma_{0,bosonic/fermionic,brane}$) tends
to infinity ($\infty$) and consequently, attractive force is removed  and  quarks and anti-quarks become free. \\

 To show this, we use the method in \cite{k24} and assume that one gauge field like one photon moves between quarks and
 anti-quarks. The wave equation for this particle is:

  \begin{eqnarray}
  && -\frac{\partial^{2} A^{i}}{\partial t^{2}} + \frac{\partial^{2}
  A^{i}}{\partial z^{2}}=0 \label{kp30}
  \end{eqnarray}

 Here, $z$ is the length of wormhole which connects quark and anti-quarks. Using the below re-parameterizations \cite{k24}:

  \begin{eqnarray}
  && \rho = \frac{z^{2}}{w} ,  \nonumber \\&& w=
  \frac{V_{tot}}{2E_{system}}\nonumber \\&&
  \bar{\tau} = \gamma\int_{0}^{t} d\tau' \frac{w}{\dot{w}} - \gamma
  \frac{z^{2}}{2}\label{kp31}
  \end{eqnarray}

  and doing the  below calculations:

  \begin{eqnarray}
  \left [\left\{(\frac{\partial \bar{\tau}}{\partial t})^{2} -
  (\frac{\partial \bar{\tau}}{\partial
  z})^{2}\right\}\frac{\partial^{2}}{\partial
  t^{2}}+\left\{(\frac{\partial \rho}{\partial z})^{2} -
  (\frac{\partial \rho}{\partial t})^{2} \right
  \}\frac{\partial^{2}}{\partial \rho^{2}}\right]X^{i}=0
  \label{kp32}
  \end{eqnarray}

  we get \cite{k24}:

  \begin{eqnarray}
  && (-g)^{-1/2}\frac{\partial}{\partial
  x_{\mu}}[(-g)^{1/2}g^{\mu\nu}]\frac{\partial}{\partial
  x_{\upsilon}}X^{i}=0\label{kp33}
  \end{eqnarray}

  where $x_{0}=\bar{\tau}$, $x_{1}=\rho$ and the line elements are
  obtained by:

  \begin{eqnarray}
  && g^{\bar{\tau}\bar{\tau}}\sim
  -\frac{1}{\beta^{2}}(\frac{w'}{w})^{2}\frac{(1-(\frac{w}{w'})^{2}\frac{1}{z^{4}})}{(1+(\frac{w}{w'})^{2}
  \frac{(1+\gamma^{-2})}{z^{4}})^{1/2}}\nonumber
  \\&&g^{\rho\rho}\sim -(g^{\bar{\tau}\bar{\tau}})^{-1}\label{kp34}
  \end{eqnarray}

At this stage, we  compare above elements  with the metric
of a thermal BIon \cite{k24}:

\begin{eqnarray}
&& ds^{2}= D^{-1/2}\bar{H}^{-1/2}(-f
dt^{2}+dx_{1}^{2})+D^{1/2}\bar{H}^{-1/2}(dx_{2}^{2}+dx_{3}^{2})+D^{-1/2}\bar{H}^{1/2}(f^{-1}
dr^{2}+r^{2}d\Omega_{5})^{2}\nonumber
\\&&\label{kp35}
\end{eqnarray}

where

\begin{eqnarray}
&&f=1-\frac{r_{0}^{4}}{r^{4}},\nonumber
\\&&\bar{H}=1+\frac{r_{0}^{4}}{r^{4}}sinh^{2}\alpha \nonumber
\\&&D^{-1}=cos^{2}\varepsilon + H^{-1}sin^{2}\varepsilon \nonumber
\\&&cos\varepsilon =\frac{1}{\sqrt{1+\frac{\beta^{2}}{\sigma^{p-1}}}}
\label{kp36}
\end{eqnarray}

Comparing the metric of (\ref{kp36}) with the  metric
of (\ref{kp34}), we derive the following relations \cite{k24}:

\begin{eqnarray}
&&f=1-\frac{r_{0}^{4}}{r^{4}}\sim
1-\left(\frac{w}{w'}\right)^{2}\frac{1}{z^{4}},\nonumber
\\&&\bar{H}=1+\frac{r_{0}^{4}}{r^{4}}sinh^{2}\alpha \sim 1+\left(\frac{w}{w'}\right)^{2}\frac{(1+\gamma^{-2})}{z^{4}} \nonumber
\\&&D^{-1}=cos^{2}\varepsilon + \bar{H}^{-1}sin^{2}\varepsilon\simeq1\nonumber
\\&&
\Rightarrow r\sim z,r_{0}\sim
\left(\frac{w}{w'}\right)^{1/2},(1+\gamma^{-2})\sim sinh^{2}\alpha
\nonumber
\\&&cosh^{2}\alpha = \frac{3}{2}\frac{cos\frac{\delta}{3} +
\sqrt{3}sin\frac{\delta}{3}}{cos\delta}\nonumber
\\&& cos\delta \equiv \overline{T}^{4}F_{Total},\, \overline{T} \equiv
\frac{T}{T_{c}} \nonumber
\\&& F_{Total}=F_{bosonic,brane,tot}+F_{fermionic,brane,tot}+F_{bosonic,anti-brane,tot}+F_{fermionic,anti-brane,tot}\label{kp37}
\end{eqnarray}

The BIonic temperature is defined by  ${\displaystyle T=\frac{1}{\pi r_{0}
cosh\alpha}}$. Consequently, the temperature of a BIon has the following relation with the potential:

\begin{eqnarray}
&& T=\frac{1}{\pi r_{0}
cosh\alpha}=\frac{\gamma}{\pi}(\frac{w'}{w})^{1/2}\sim
\frac{\gamma}{\pi}(\frac{V'_{tot}}{V_{tot}})^{1/2}\sim \nonumber
\\&&\frac{\gamma}{\pi}[ -\Sigma_{m=1}^{P-1}\Sigma_{n=1}^{N-1} [k_{bosonic,brane,}-k_{bosonic,anti-brane}]^{m}m\sigma^{m-1}
\Big(\sigma_{0,bosonic,brane}^{nm}-\sigma^{nm}\Big)+\nonumber
\\&&\Sigma_{m=1}^{P-1}\Sigma_{n=1}^{N-1} [k_{bosonic,brane,}-k_{bosonic,anti-brane}]^{m}nm\sigma^{m}\Big(\sigma^{nm-1}\Big)-\nonumber
\\&& \Sigma_{m=1}^{P-1}\Sigma_{n=1}^{N-1} [k_{fermionic,brane,}-k_{fermionic,anti-brane}]^{m}\frac{m}{\sigma^{m+1}}
\Big(\sigma^{-nm}-\sigma^{-nm}_{0,fermionic,brane}\Big)-\nonumber
\\&& \Sigma_{m=1}^{P-1}\Sigma_{n=1}^{N-1} [k_{fermionic,brane,}-k_{fermionic,anti-brane}]^{m}\frac{nm}{\sigma^{m}}
\Big(\sigma^{-nm-1}\Big)]^{1/2}\times\nonumber
\\&&[- \Sigma_{m=1}^{P-1}\Sigma_{n=1}^{N-1} [k_{bosonic,brane,}-k_{bosonic,anti-brane}]^{m}\sigma^{m}
\Big(\sigma_{0,bosonic,brane}^{nm}-\sigma^{nm}\Big)+\nonumber
\\&& \Sigma_{m=1}^{P-1}\Sigma_{n=1}^{N-1} [k_{fermionic,brane,}-k_{fermionic,anti-brane}]^{m}\frac{1}{\sigma^{m}}
\Big(\sigma^{-nm}-\sigma^{-nm}_{0,fermionic,brane}\Big)]^{-1/2} \label{kp38}
\end{eqnarray}

From this point of view that $\gamma$ has the relation the temperature, we get \cite{k24}:

\begin{eqnarray}
&&\gamma = \frac{1}{cosh\alpha} \sim
\frac{2cos\delta}{3\sqrt{3}-cos\delta
-\frac{\sqrt{3}}{6}cos^{2}\delta}\sim \nonumber
\\&&\frac{2\overline{T}^{4}F_{Total}}{3\sqrt{3}-\overline{T}^{4}F_{Total}
-\frac{\sqrt{3}}{6}\overline{T}^{8}F_{Total}^{2}}
\label{kp39}
\end{eqnarray}

To similarity, we assume that throats of bosonic and fermionic worhomes have the same size
($\sigma_{0,bosonic,brane}=\sigma_{0,fermionic,brane}=\sigma_{0,bosonic,anti-brane}=\sigma_{0fermionic,anti-brane}$) and
also  ($k_{bosonic,brane}-k_{bosonic,anti-brane}=k_{fermionic, brane}-k_{fermionic,anti-brane}$). Using Eqs
(\ref{kp37}, \ref{kp38} and \ref{kp39}), we can obtain the approximate form of the separation distance between quarks
and anti-quarks in terms of temperature:

\begin{eqnarray}
&& \sigma \approx \Sigma_{m=1}^{P-1}\Sigma_{n=1}^{N-1} [k_{bosonic,brane,}-k_{bosonic,anti-brane}]^{-m}
[(\frac{\sqrt{3}}{6}\overline{T}^{2}T)^{\frac{1}{nm+m+1}}+(\frac{T}{2})^{\frac{2}{nm+m+1}}]\nonumber
\\&& \sigma_{0,bosonic,brane}=\sigma_{0,fermionic,brane}=\Sigma_{m=1}^{P-1}\Sigma_{n=1}^{N-1}
[k_{bosonic,brane,}-k_{bosonic,anti-brane}]^{-m-nm}\times \nonumber
\\&&[(\frac{\sqrt{3}}{6}\overline{T}^{2}T)^{\frac{1}{nm+m+1}}+(\frac{T}{2})^{\frac{2}{nm+m+1}}]^{\frac{1+nm}{nm}}
[\frac{\overline{T}^{4}+\frac{\sqrt{3}}{6}\overline{T}^{8}}{2\overline{T}^{4}}]^{\frac{1}{nm+1}}[\frac{T}{T_{c}-T}]^{\frac{2}{nm+1}}
\label{kp40}
\end{eqnarray}

This equation shows that by increasing temperature, the place of  boundary between repulsive and attractive force
($\sigma_{0,bosonic/fermionic,brane/anti-brane}$) changes and goes to infinity at a critical temperature and thus,
quarks and anti-quarks become free at this point. This result is in agreement with experiments. In fact, by increasing
temperature, energy of particles increases and they can overcome attractive force and deconfinement emerges. We will
demonstrate this by calculating the bosinc and fermionic potentials in terms of temperature. The relation between entropies and
potentials are as follows \cite{k6,k23}:

\begin{eqnarray}
&& F_{tot}=M_{tot}-T\bar{S}
\label{kpp40}
\end{eqnarray}

where $F_{tot}$ is the free energy for this system which has direct relation with Hamiltonian and potential
( $F_{tot}\approx V_{tot}$). Also, $M_{tot}$ is total mass of system which is related to total energy of system
$E_{tot}=M_{tot}$, $T$ is temperature and $\bar{S}$ is entropy. In this mechanism, all things are produced  from
nothing as discussed already and after Eq. (\ref{t1}) and thus $E_{tot}$ is zero and $\bar{S}\approx \frac{V_{tot}}{T}$.
Substituting Eq. (\ref{kp40}) in Eq. (\ref{fff1}), we obtain potentials and entropies as:

      \begin{eqnarray}
     &&   V_{tot}=V_{bosonic,brane+anti-brane,tot}+V_{fermionic,brane+anti-brane,tot}\nonumber\\&&\nonumber\\&&\nonumber\\&&
     \bar{S}_{tot}=\bar{S}_{bosonic,brane+anti-brane,tot}+\bar{S}_{fermionic,brane+anti-brane,tot}\approx\nonumber\\&&-
     \frac{V_{tot}}{T}=\frac{V_{bosonic,brane+anti-brane,tot}+V_{fermionic,brane+anti-brane,tot}}{T}\label{k1fp41}
     \end{eqnarray}

  \begin{eqnarray}
  &&   \bar{S}_{bosonic,brane+anti-brane,tot}\approx -\frac{V_{bosonic,brane+anti-brane,tot}}{T}\approx
     \nonumber\\&& \frac{1}{T}\Sigma_{m=1}^{P-1}\Sigma_{n=1}^{N-1} [k_{bosonic,brane,}-k_{bosonic,anti-brane}]^{m}\times
     \nonumber\\&&[\Sigma_{m=1}^{P-1}\Sigma_{n=1}^{N-1} [k_{bosonic,brane,}-k_{bosonic,anti-brane}]^{-m}[(\frac{\sqrt{3}}{6}
     \overline{T}^{2}T)^{\frac{1}{nm+m+1}}+(\frac{T}{2})^{\frac{2}{nm+m+1}}]]^{m}\times\nonumber\\&&\Big([\Sigma_{m=1}^{P-1}
     \Sigma_{n=1}^{N-1} [k_{bosonic,brane,}-k_{bosonic,anti-brane}]^{-m-nm}\times \nonumber
      \\&&[(\frac{\sqrt{3}}{6}\overline{T}^{2}T)^{\frac{1}{nm+m+1}}+(\frac{T}{2})^{\frac{2}{nm+m+1}}]^{\frac{1+nm}{nm}}
      [\frac{\overline{T}^{4}+\frac{\sqrt{3}}{6}\overline{T}^{8}}{2\overline{T}^{4}}]^{\frac{1}{nm+1}}[\frac{T}{T_{c}-T}]^{\frac{2}
      {nm+1}}]^{nm}-\nonumber\\&&[\Sigma_{m=1}^{P-1}\Sigma_{n=1}^{N-1} [k_{bosonic,brane,}-k_{bosonic,anti-brane}]^{-m}
      [(\frac{\sqrt{3}}{6}\overline{T}^{2}T)^{\frac{1}{nm+m+1}}+(\frac{T}{2})^{\frac{2}{nm+m+1}}]]^{nm}\Big)\Rightarrow\label{o2k3fp41}
      \end{eqnarray}
      \begin{eqnarray}
      &&  F_{bosonic,repulsive}\approx\Sigma_{m=1}^{P-1}\Sigma_{n=1}^{N-1} [k_{bosonic,brane,}-
      k_{bosonic,anti-brane}]^{m}m\times \nonumber\\&&[\Sigma_{m=1}^{P-1}\Sigma_{n=1}^{N-1} [k_{bosonic,brane,}-
      k_{bosonic,anti-brane}]^{-m}[(\frac{\sqrt{3}}{6}\overline{T}^{2}T)^{\frac{1}{nm+m+1}}+(\frac{T}{2})^{\frac{2}{nm+m+1}}]]^{m-1}
      \times\nonumber\\&&[\Sigma_{m=1}^{P-1}\Sigma_{n=1}^{N-1} [k_{bosonic,brane,}-k_{bosonic,anti-brane}]^{-m-nm}\times \nonumber
      \\&&[(\frac{\sqrt{3}}{6}\overline{T}^{2}T)^{\frac{1}{nm+m+1}}+(\frac{T}{2})^{\frac{2}{nm+m+1}}]^{\frac{1+nm}{nm}}
      [\frac{\overline{T}^{4}+\frac{\sqrt{3}}{6}\overline{T}^{8}}{2\overline{T}^{4}}]^{\frac{1}{nm+1}}
      [\frac{T}{T_{c}-T}]^{\frac{2}{nm+1}}]^{nm}\label{o1k3fp41}
      \end{eqnarray}
      \begin{eqnarray}
      &&
      F_{bosonic,attractive}\approx-\Sigma_{m=1}^{P-1}\Sigma_{n=1}^{N-1} [k_{bosonic,brane,}-k_{bosonic,anti-brane}]^{m}n(m+1)
      \times \nonumber\\&&[\Sigma_{m=1}^{P-1}\Sigma_{n=1}^{N-1} [k_{bosonic,brane,}-k_{bosonic,anti-brane}]^{-m}
      [(\frac{\sqrt{3}}{6}\overline{T}^{2}T)^{\frac{1}{nm+m+1}}+(\frac{T}{2})^{\frac{2}{nm+m+1}}]]^{m+nm-1}\label{k2fp41}
      \end{eqnarray}
    \begin{eqnarray}
      &&\bar{S}_{fermionic,brane+anti-brane,tot}\approx -\frac{V_{fermionic,brane+anti-brane,tot}}{T}\approx \nonumber\\&& -
      \frac{1}{T}\Sigma_{m=1}^{P-1}\Sigma_{n=1}^{N-1} [k_{fermionic,brane,}-k_{fermionic,anti-brane}]^{m}\times\nonumber\\&&
      [\Sigma_{m=1}^{P-1}\Sigma_{n=1}^{N-1} [k_{bosonic,brane,}-k_{bosonic,anti-brane}]^{-m}[(\frac{\sqrt{3}}{6}
      \overline{T}^{2}T)^{\frac{1}{nm+m+1}}+(\frac{T}{2})^{\frac{2}{nm+m+1}}]]^{-m}\times\nonumber\\&&\Big([\Sigma_{m=1}^{P-1}
      \Sigma_{n=1}^{N-1} [k_{bosonic,brane,}-k_{bosonic,anti-brane}]^{-m}[(\frac{\sqrt{3}}{6}\overline{T}^{2}T)^{\frac{1}{nm+m+1}}+
      (\frac{T}{2})^{\frac{2}{nm+m+1}}]]^{-nm}-\nonumber\\&&[\Sigma_{m=1}^{P-1}\Sigma_{n=1}^{N-1} [k_{bosonic,brane,}-
      k_{bosonic,anti-brane}]^{-m-nm}\times \nonumber
       \\&&[(\frac{\sqrt{3}}{6}\overline{T}^{2}T)^{\frac{1}{nm+m+1}}+(\frac{T}{2})^{\frac{2}{nm+m+1}}]^{\frac{1+nm}{nm}}
       [\frac{\overline{T}^{4}+\frac{\sqrt{3}}{6}\overline{T}^{8}}{2\overline{T}^{4}}]^{\frac{1}{nm+1}}
       [\frac{T}{T_{c}-T}]^{\frac{2}{nm+1}}]^{-nm}\Big) \Rightarrow \label{s2k3fp41}
       \end{eqnarray}
       \begin{eqnarray}
       &&
       F_{fermionic,repulsive}\approx \Sigma_{m=1}^{P-1}\Sigma_{n=1}^{N-1} [k_{fermionic,brane,}-k_{fermionic,anti-brane}]^{m}(m+nm)
       \times \nonumber\\&&[\Sigma_{m=1}^{P-1}\Sigma_{n=1}^{N-1} [k_{bosonic,brane,}-k_{bosonic,anti-brane}]^{-m}
       [(\frac{\sqrt{3}}{6}\overline{T}^{2}T)^{\frac{1}{nm+m+1}}+(\frac{T}{2})^{\frac{2}{nm+m+1}}]]^{-m-nm-1}\label{s1k3fp41}
       \end{eqnarray}
   \begin{eqnarray}
   &&
       F_{fermionic, attractive}\approx-\Sigma_{m=1}^{P-1}\Sigma_{n=1}^{N-1} [k_{fermionic,brane,}-k_{fermionic,anti-brane}]^{m}
        (m+1)\times \nonumber\\&&[\Sigma_{m=1}^{P-1}\Sigma_{n=1}^{N-1} [k_{bosonic,brane,}-k_{bosonic,anti-brane}]^{-m}[(\frac{\sqrt{3}}
        {6}\overline{T}^{2}T)^{\frac{1}{nm+m+1}}+(\frac{T}{2})^{\frac{2}{nm+m+1}}]]^{-m-1}\times \nonumber\\&&[\Sigma_{m=1}^{P-1}
        \Sigma_{n=1}^{N-1} [k_{bosonic,brane,}-k_{bosonic,anti-brane}]^{-m-nm}\times \nonumber
     \\&&[(\frac{\sqrt{3}}{6}\overline{T}^{2}T)^{\frac{1}{nm+m+1}}+(\frac{T}{2})^{\frac{2}{nm+m+1}}]^{\frac{1+nm}{nm}}
     [\frac{\overline{T}^{4}+\frac{\sqrt{3}}{6}\overline{T}^{8}}{2\overline{T}^{4}}]^{\frac{1}{nm+1}}
     [\frac{T}{T_{c}-T}]^{\frac{2}{nm+1}}]^{-nm}  \label{k3fp41}
     \end{eqnarray}

 These results show that bosonic  entropy is zero near $T=0$, which grows with temperature and tends to infinity at $\infty$.
 This entropy includes two types of terms$-$some with positive sign and some with negative sign. Terms with positive sign
 produce repulsive force and those with negative sign create the attractive force. With increasing temperature, repulsive
 terms grow and tend to $\infty$ at $T=T_{c}$, while, attractive terms increase with lower velocity. On the other hand,
 the fermionic  entropy is infinity near $T=0$, decrease and shrink to zero at higher temperatures. This entropy also consists
 of negative terms which reverse to bosonic one, produce repulsive force  and positive terms which reverse to the case of
 bosonic wormhole, create the attractive force. The attractive terms decrease faster than repulsive terms and shrink to zero
 at $T=T_{c}$. Thus, the repulsive force which is produced by both fermionic and bosonic wormholes overcomes attractive
 force at this point and produces deconfinement. These results are in good agreement with experimental data and previous
 predictions from QCD in \cite{k1,k2,k3,k4,k5,k6}. All above dependences  are describe on Figures 1-4. From figure 3, it is clear that for low temperature the attractive
force between bosonic states of quarks are more than repelling force,
while in figure 4, repulsive force for fermionic states of quarks is
more than attractive force. If we sum over these forces, we observe
that for low temperature quarks repel each other. This is in agreement
with previous prediction of QCD that in a quarkonium quarks can't
become very close to each other.
From figure 3, it is also clear that for high temperature the repulsive
force between bosonic states of quarks are more than attractive force,
while in figure 4, attractive force for fermionic states of quarks is
more than repulsive force. If we sum over these forces, we observe
that for low temperature quarks attract each other. This is in
agreement with previous prediction of QCD that in a quarkonium quarks
can't become very distant from each other.
Thus, in our model, totally quarks can't become very close or very
distant from each other and are approximately free in middle distant
in quarkonium. This is in agreement with QCD.
%%%%%%%%%%%%%%%%%%%%%%%%%%%%%%%%%%%%%%%%%%%%%% Figure 1 %%%%%%%%%%%%%%%%%%%%%%%%%%%%%%%%%%%%%%%%%%%%%%%
\begin{figure*}[thbp]
 	\begin{center}
	 	\includegraphics[width=.45\textwidth]{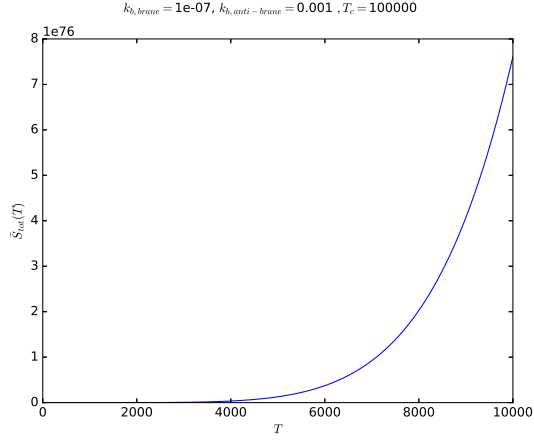}
 	\end{center}
 	\caption{The total entropy $\bar{S}_{tot}$ from Eq.~(\ref{k1fp41}).}
 \end{figure*}
%%%%%%%%%%%%%%%%%%%%%%%%%%%%%%%%%%%%%%%%%%%%%%%%%%%%%%%%%%%%%%%%%%%%%%%%%%%%%%%%%%%%%%%%%%%%%%%%%%%%%%%

%%%%%%%%%%%%%%%%%%%%%%%%%%%%%%%%%%%%%%%%%%%%%% Figure 2 %%%%%%%%%%%%%%%%%%%%%%%%%%%%%%%%%%%%%%%%%%%%%%%
\begin{figure*}[thbp]
	\begin{center}
		\includegraphics[width=.45\textwidth]{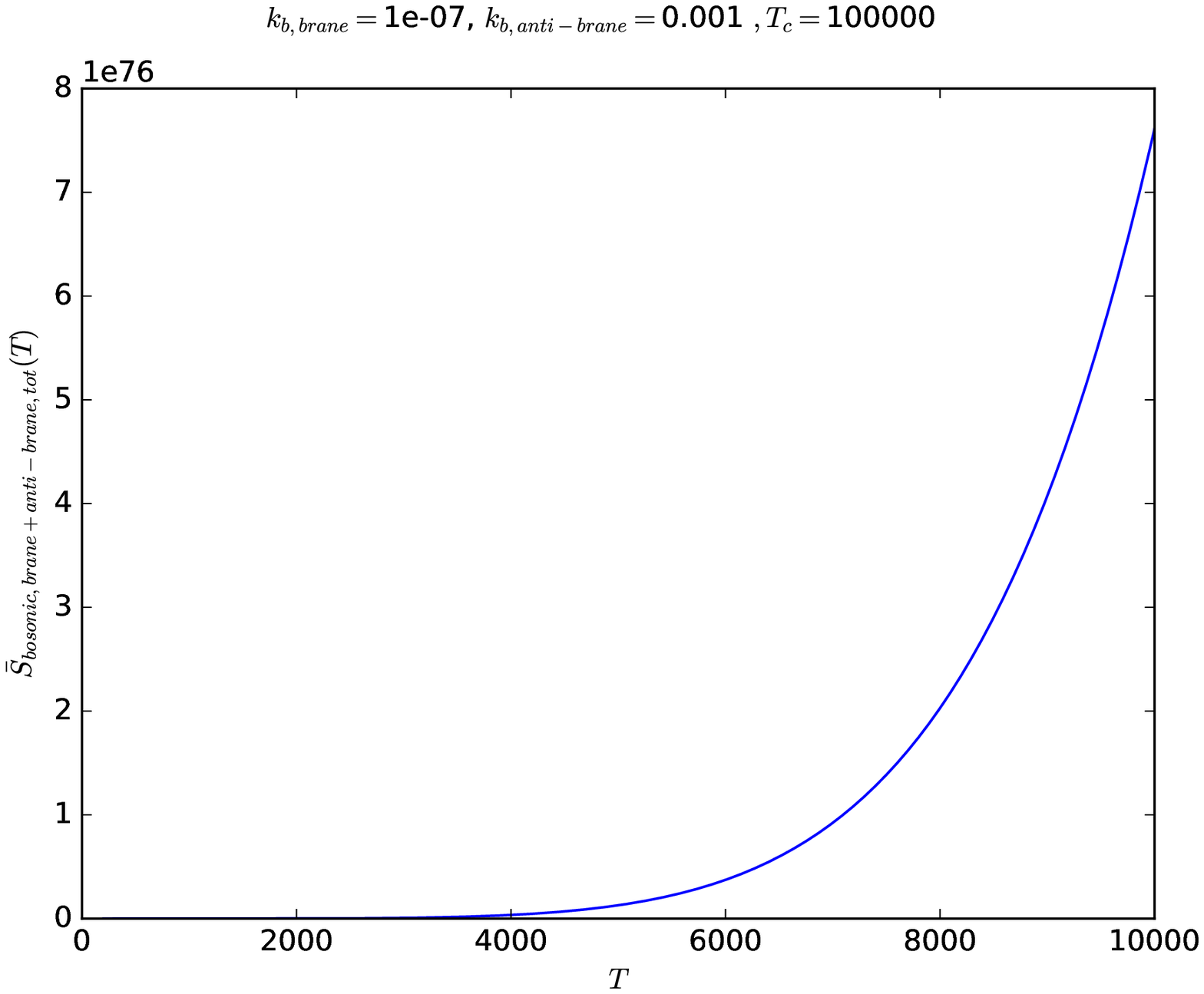}\hspace{2em}
		\includegraphics[width=.45\textwidth]{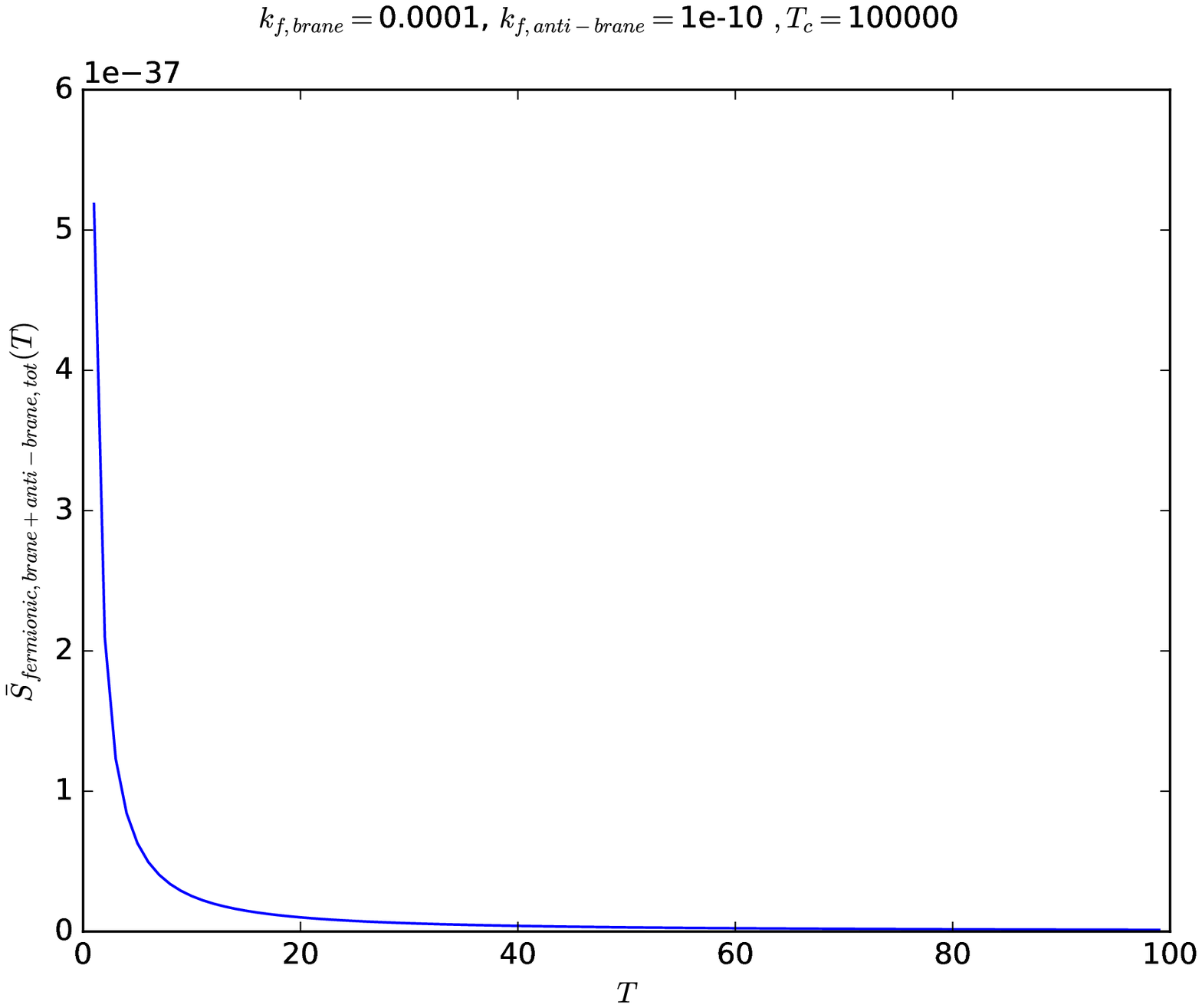}
	\end{center}
	\caption{(left) The bosonic entropy $\bar{S}_{bosonic,brane+anti-brane,tot}$ from Eq.~(\ref{o2k3fp41}); (right)
	the fermionic entropy $\bar{S}_{fermionic,brane+anti-brane,tot}$ from Eq.~(\ref{s2k3fp41}).}
\end{figure*}
%%%%%%%%%%%%%%%%%%%%%%%%%%%%%%%%%%%%%%%%%%%%%%%%%%%%%%%%%%%%%%%%%%%%%%%%%%%%%%%%%%%%%%%%%%%%%%%%%%%%%%%

%%%%%%%%%%%%%%%%%%%%%%%%%%%%%%%%%%%%%%%%%%%%%% Figure 3 %%%%%%%%%%%%%%%%%%%%%%%%%%%%%%%%%%%%%%%%%%%%%%%
\begin{figure*}[thbp]
	\begin{center}
		\includegraphics[width=.45\textwidth]{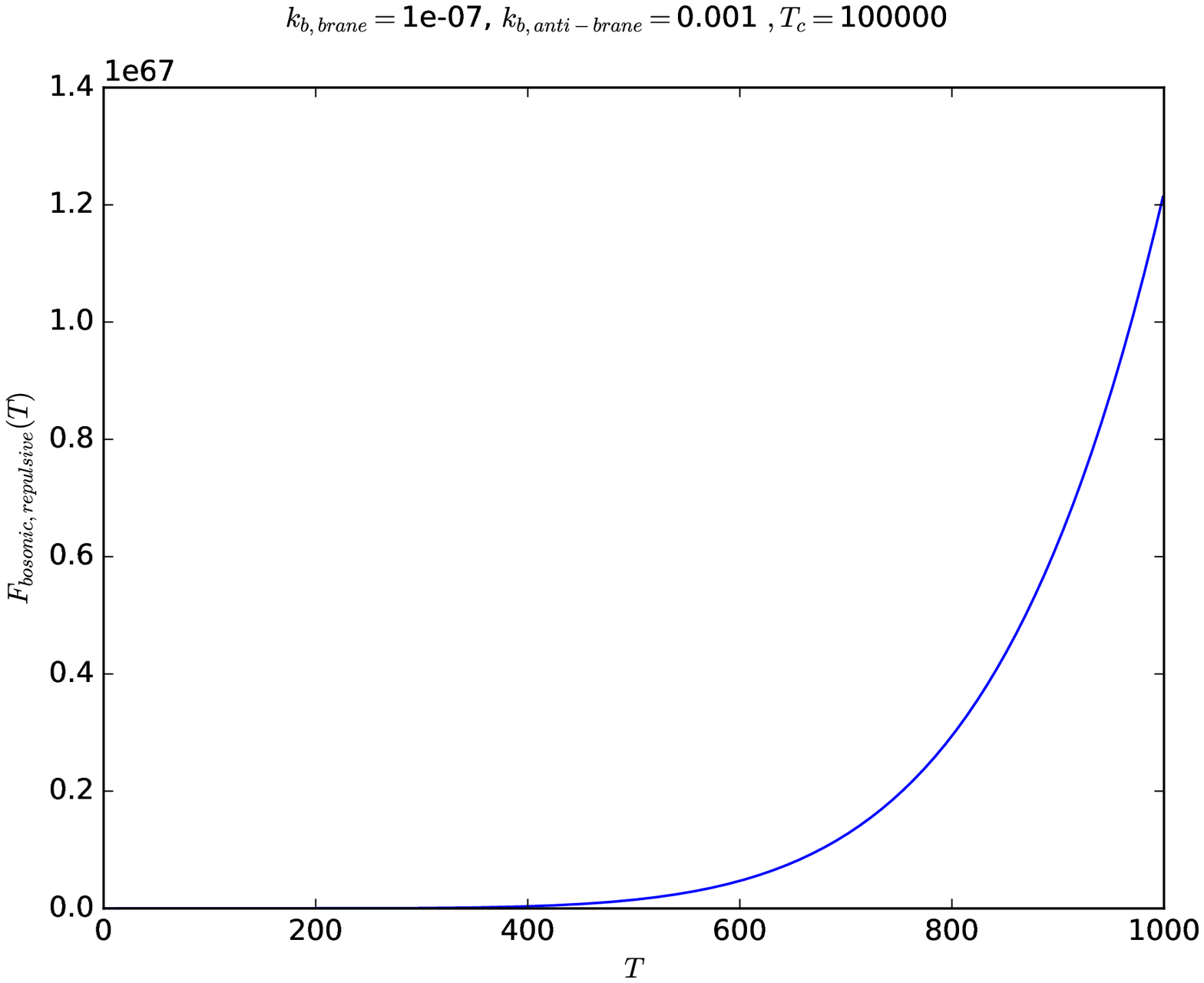}\hspace{2em}
		\includegraphics[width=.45\textwidth]{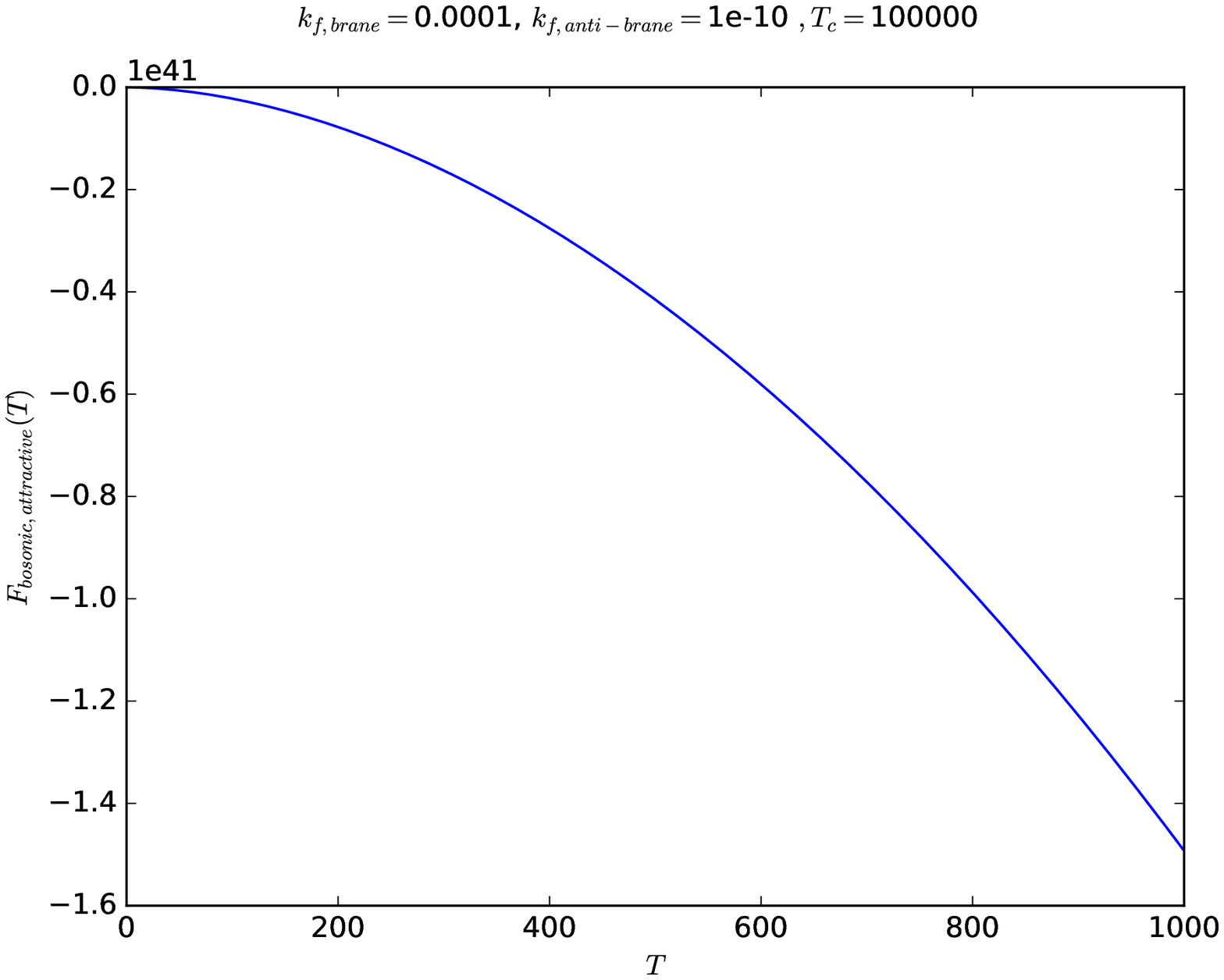}
	\end{center}
	\caption{(left) The repulsive force $F_{bosonic,repulsive}$ from Eq.~(\ref{o1k3fp41}); (right) the attractive
	force $\bar{F}_{bosonic,attractive}$ from Eq.~(\ref{k2fp41}).}
\end{figure*}
%%%%%%%%%%%%%%%%%%%%%%%%%%%%%%%%%%%%%%%%%%%%%%%%%%%%%%%%%%%%%%%%%%%%%%%%%%%%%%%%%%%%%%%%%%%%%%%%%%%%%%%

%%%%%%%%%%%%%%%%%%%%%%%%%%%%%%%%%%%%%%%%%%%%%% Figure 4 %%%%%%%%%%%%%%%%%%%%%%%%%%%%%%%%%%%%%%%%%%%%%%%
\begin{figure*}[thbp]
	\begin{center}
		\includegraphics[width=.45\textwidth]{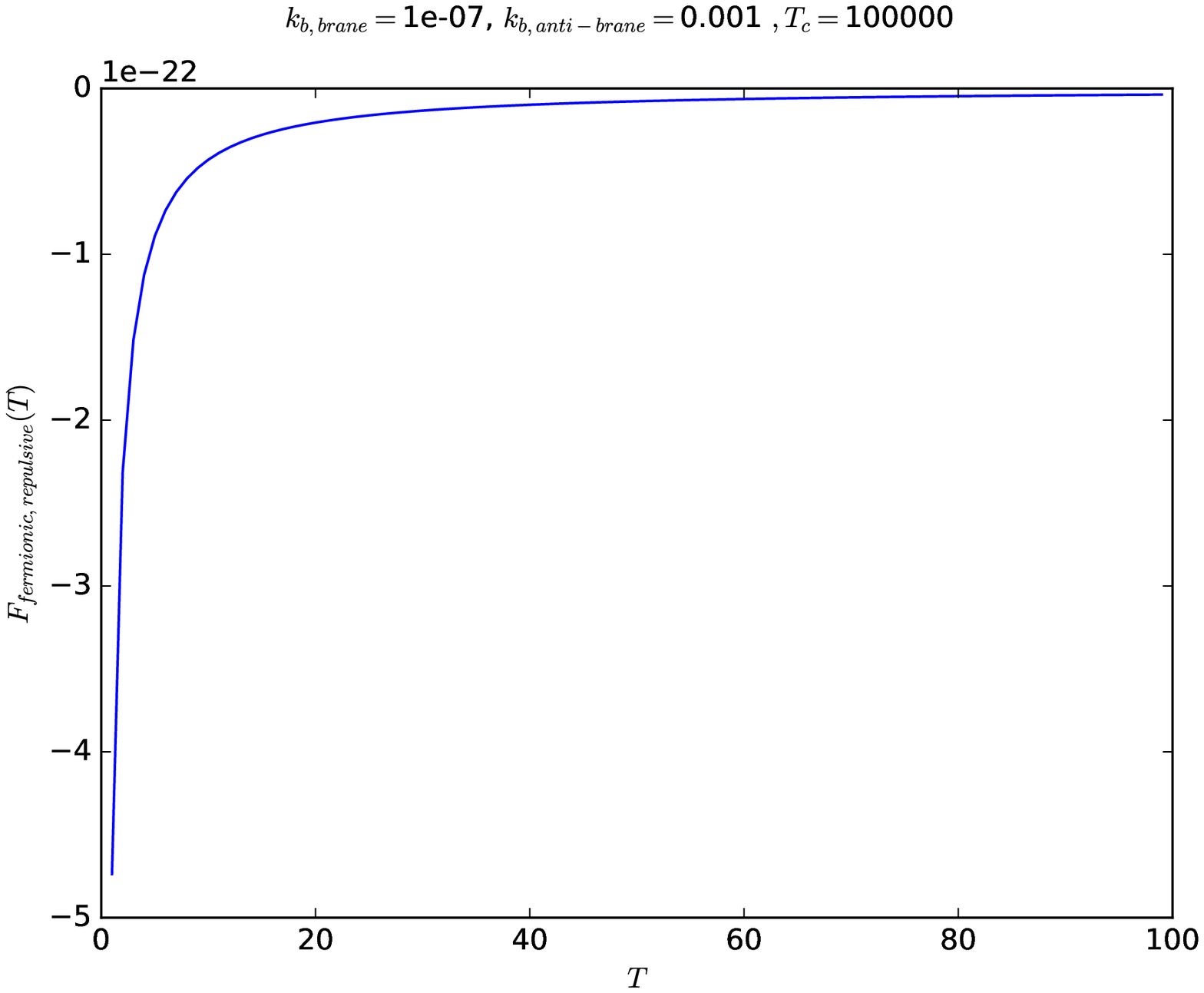}\hspace{2em}
		\includegraphics[width=.45\textwidth]{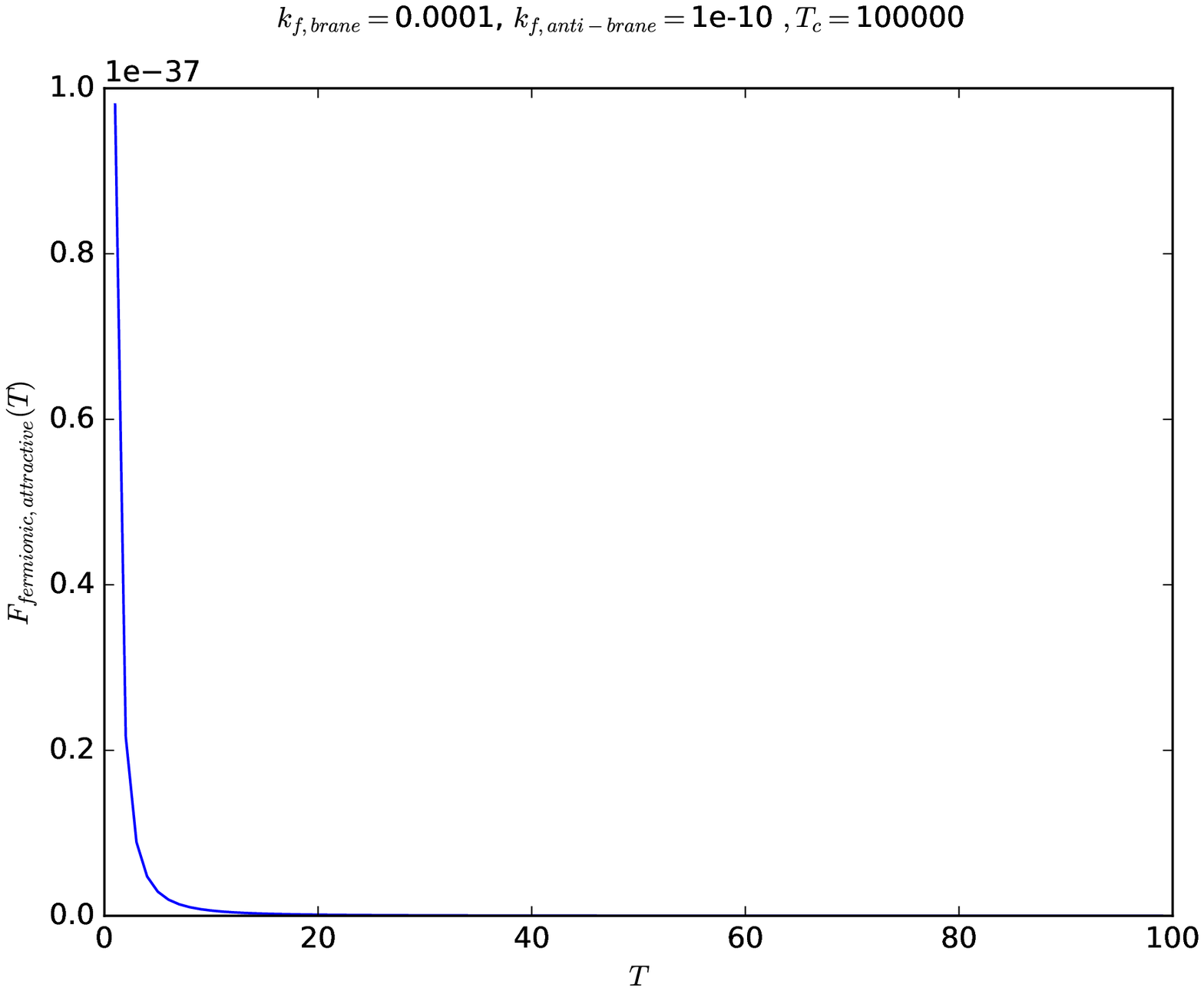}
	\end{center}
	\caption{(left) The repulsive force $F_{fermionic,repulsive}$ from Eq.~(\ref{s1k3fp41}); (right) the attractive
	force $\bar{F}_{fermionic,attractive}$ from Eq.~(\ref{k3fp41}).}
\end{figure*}
%%%%%%%%%%%%%%%%%%%%%%%%%%%%%%%%%%%%%%%%%%%%%%%%%%%%%%%%%%%%%%%%%%%%%%%%%%%%%%%%%%%%%%%%%%%%%%%%%%%%%%%

In figure 5, we have obtained the bosonic and the fermionic potentials in terms of temperature for N=3, P=4 and $T_{c}=.1 GeV$  .  It is clear that when temperature of a quark and an anti-quark becomes zero in a quarkonium, they meet each other and    both bosonic and fermionic potentials become infinite. In the middle temperature that quarks and anti-quarks are separated approximately. The negative bosonic potential  and bosonic fermionic potential cancel the effect of each other and there existed a freedom like the same as  predicted in QCD. By achieving temperature to critical temperature ($.1-.3 GeV$), total potential becomes zero and a real deconfinement is appeared. These results are in agreement with previous prediction for deconfinement in \cite{B1,B2,B3}. By increasing temperature, two quarks become far from each other and both potentials grow and tend to large positive values.  In these conditions, the deconfinement of system increases  as can be seen in the energies of LHC (See for example \cite{B4} ). Thus, our model gives true value for critical temperature which really has been seen in experiments.

%%%%%%%%%%%%%%%%%%%%%%%%%%%%%%%%%%%%%%%%%%%%%% Figure 1 %%%%%%%%%%%%%%%%%%%%%%%%%%%%%%%%%%%%%%%%%%%%%%%
\begin{figure*}[thbp]
	\begin{center}
		\includegraphics[width=.45\textwidth]{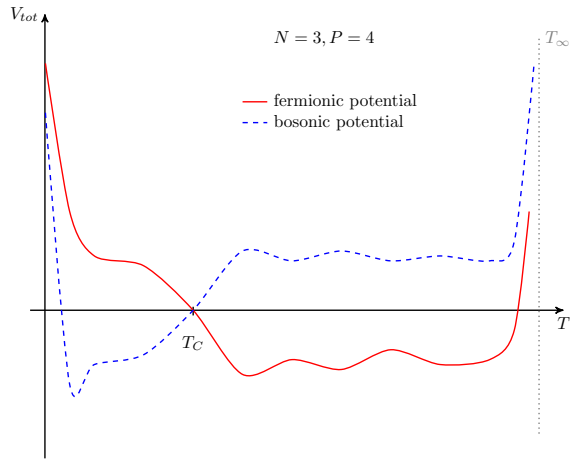}
	\end{center}
	\caption{The bosonic and fermionic potential between quark and anti-quark for N=3, P=4 and $T_{c}=.1 GeV$ .}
\end{figure*}
%%%%%%%%%%%%%%%%%%%%%%%%%%%%%%%%%%%%%%%%%%%%%%%%%%%%%%%%%%%%%%%%%%%%%%%%%%%%%%%%%%%%%%%%%%%%%%%%%%%%%%%

%%%%%%%%%%%%%%%%%%%%%%%%%%%%%%%%%%%%%%%%%%%%%%%%%%%%%%%%%%%%%%%%%%%%%%%%%%%%%%%%%%%%%%%%%%%%%%%%%%%%%
%%%%%%%%%%%%%%%%%%%%%%%%%%%%%%%%%%%% SECTION III %%%%%%%%%%%%%%%%%%%%%%%%%%%%%%%%%%%%%%%%%%%%%%%%
\pagebreak

\section{Summary and conclusion }\label{o4}
In this paper, we have considered the process of birth of quarks, anti-quarks and confiding potential between
them which leads to formation of quarkonium in a thermal BIon. Quarkonium is constructed of one quark and one
anti-quark that are confined to each other and can't become very close to each other or go much away from each other.
By closing quarks to anti-quarks, they are paired and form an scalar system. However, by getting away of these particles,
the fermionic properties overcome. Thus, we need a theory that fermions and bosons have the same origin and transit
to each other in it. In $M$-theory, these two types of particles are completely independent and for this reason, we
introduce BLNA-theory that has higher dimensions respect to $M$-theory and is reduced to it in $11$-dimensions. In this theory,
at the beginning, there is no degree of freedom and energy. Then, two types of energies emerge that are only
different in their sign and sum  over them is zero. Each of these energies creates some degrees of freedom which lead
to production of two types of branes with opposite quantum numbers. Coinciding with the birth of these branes, some
bosonic tensor fields are born with their rank is changed from zero to dimension of brane and appear as scalar fields
and gravitons in four dimensions. These fields  interact with fields of other branes and cause the branes to be compacted.
By compacting of branes, fermions emerge which some of them with lower spins play the role of quarks and anti-quarks
and some other with higher spins have the role of gravitino. In this system, gravitons create a bosonic wormhole that leads
to attractive potential in large separation distance between quarks and anti-quarks and prevents them from getting away
from each other. Also, gravitinos produce fermionic wormhole that causes to the emergence of repulsive force in small
separation distance between quarks and anti-quarks and prevents them from closing into each other. The confiding potential
which is produced by these wormholes can be reduced to previous predicted potential in QCD and experimental data. With
increasing temperature, these two wormholes produce two types of entropies which lead to the emergence of repulsive force
at higher temperatures. The bosonic entropy is zero near $T=0$, which grows with temperature and tends to infinity at $\infty$.
This entropy contains two types of terms, positive terms which  create repulsive force and negative terms which produce the
attractive force. With increasing temperature, repulsive terms grow and tend to $\infty$ at $T=T_{c}$, while, attractive
terms increase with lower velocity.  Also, the fermionic  entropy is infinity near $T=0$, decreases and shrinks to zero at
higher temperatures. This entropy also includes  negative terms which reverse to bosonic one, produces repulsive force
and positive terms which reverse to the case of bosonic wormhole, create the attractive force. The attractive terms
decreases faster than repulsive terms and shrinks to zero at $T=T_{c}$. Thus, total entropy produces repulsive force which
overcomes to attractive force in higher temperature and leads to the deconfinement.

%%%%%%%%%%%%%%%%%%%%%%%%%%%%%%%%%%%%%%%%%%%%%%%%%%%%%%%%%%%%%%%%%%%%%%%%%%%%%%%%%%%%%%%%%
\section*{Acknowledgments}
\noindent
The work of Alireza Sepehri has been supported financially by Research
Institute for Astronomy-Astrophysics of Maragha (RIAAM) under research
project NO.1/4717-92.
 The work was partly supported by
VEGA Grant No. 2/0009/16 and No. 1/0222/13 of the Ministry of Education, Science, Research and Sport of the
Slovak Republic. R. Pincak would like to thank the TH division in CERN for hospitality. The authors gratefully
acknowledge the hospitality of the Bogoliubov Laboratory of Theoretical Physics of the
Joint Institute for Nuclear Research, Dubna, Russian Federation. F. Rahaman and A. Pradhan thank the IUCAA, Pune, India
for providing facility and support. The authors thank the anonymous referee for fruitful comments which improved the
paper in the present form.
%%%%%%%%%%%%%%%%%%%%%%%%%%%%%%%%%%%%%%%%%%%%%%%%%%%%%%%%%%%%%%%%%%%%%%%%%%%%%%%%%%%%

\end{document}